%% file: main.tex
\documentclass[10pt,letterpaper]{article}
\usepackage[letterpaper,margin=1in]{geometry}
\usepackage[T1]{fontenc}
\usepackage{autoalpha-preprint}
\usepackage{microtype}
\usepackage{graphicx}
\usepackage{booktabs,tabularx,array}
\usepackage{xcolor}
\usepackage{autoalpha-brand}
\usepackage{listings}
\usepackage{amsmath}
\usepackage[round,authoryear]{natbib}
\usepackage{caption}
\usepackage{placeins}
\usepackage[colorlinks=true,linkcolor=AABrand,citecolor=AABrand,urlcolor=AABrand]{hyperref}
\hypersetup{pdftitle={Auto-Alpha: A Framework for Autonomous Factor Research},pdfauthor={Hui Cai, Kangcheng Deng, Tianyi Ma, Qixuan Yang, Jiacheng Lu, Rui Sun, Beidi Luan, Jing Li, Zuo Bai},pdfsubject={Methodology preprint draft}}
\title{Auto-Alpha:\\A Framework for Autonomous Factor Research}
\author{Hui Cai, Kangcheng Deng, Tianyi Ma, Qixuan Yang\\[4pt]Jiacheng Lu, Rui Sun, Beidi Luan, Jing Li, Zuo Bai\textsuperscript{*}\\[4pt]\normalfont StepFun\\[6pt]\small * Corresponding author.\\[2pt]\small Email: \href{mailto:baizuo@stepfun.com}{\texttt{baizuo@stepfun.com}}}
\date{}
\begin{document}
\maketitle
\input{sections/abstract}
\input{sections/introduction}
\input{sections/related_work}
\input{sections/setting}
\input{sections/approach}
\input{sections/empirical}
\FloatBarrier
\input{sections/insights}
\input{sections/discussion}
\input{sections/conclusion}
\bibliographystyle{plainnat}
\begingroup
\raggedright
\bibliography{references}
\endgroup
\FloatBarrier
\appendix
\input{sections/data_setup}

\FloatBarrier
\input{sections/appendix}
\FloatBarrier
\input{sections/real_case}
\clearpage
\input{sections/experience_examples}
\clearpage
\input{sections/factor_diagnostics}
\clearpage
\input{sections/phase_quality}
\clearpage
\input{sections/production_accounting}
\clearpage
\input{sections/local_production}
\end{document}

%% file: sections/abstract.tex
\begin{abstract}
\noindent Autonomous factor research requires generating economic hypotheses as well as implementing and testing them. We present Auto-Alpha, a framework that connects report-guided initialization to self-directed research through a test-driven implementation loop. An idea agent proposes economic mechanisms and signal constructions; a research agent develops them with backtest feedback. Semantic audit, quantitative review, and human admission select factors for a persistent library, while an Experience Library retains research insights. Starting from 120 research reports, Auto-Alpha builds an initial library of 32 factors. It then generates new ideas and refines existing candidates, producing 420 implementations and retaining 62 additional factors. We evaluate factor quality and portfolio performance using a fixed library snapshot of 105 factors. The median 1-day Rank IC is 0.0301, and a weekly Top-300 portfolio returns 16.81\% annualized after stated costs over 2020--2026, versus 1.71\% for the CSI 1000 price index. These retrospective results document an operating research process and its artifacts; controlled evaluation is needed to isolate the contribution of its components.
\end{abstract}

%% file: sections/introduction.tex
\section{Introduction}
\label{sec:intro}
Autonomous factor research must decide what to investigate as well as how to implement it. A useful factor idea connects an economic mechanism to observable data, a stock-ranking rule, and a test that could support or reject the proposed construction. The agent must then turn the idea into executable code, interpret backtest feedback, and revise or abandon the construction. This sequence extends the study of tool-using research agents to a concrete financial artifact \citep{lu2024aiscientist,garikaparthi2026researchgym}.

Quantitative research reports provide a practical starting point for this process. They supply mechanisms, observable proxies, comparison groups, and time scales that can guide idea generation without restricting the agent to a fixed formula grammar. Our design motivation is that such structure can make the initial proposals more useful to investigate. We use these reports to initialize a workflow that can subsequently generate its own research directions.

Auto-Alpha organizes research into report-guided bootstrapping followed by self-directed factor research. In the first stage, report-grounded ideas enter a loop of implementation, backtesting, and revision. Candidate assessment combines LLM audit, quantitative checks, and human admission. The resulting Factor Library retains executable signals, while the Experience Library records insights from successful and unsuccessful investigations. In the second stage, the idea agent proposes further research using these accumulated resources, and the same loop develops and assesses the proposals. 

We call the resulting process \emph{test-driven search scaling}. Test-driven research connects each proposed implementation or revision to diagnostic feedback. Search depth expands investigation within an idea; search breadth expands the number and variety of ideas investigated. The harness supplies tools and records while leaving research decisions to the agent. Concurrency determines how trajectories are scheduled, while cumulative attempts measure the breadth explored. Our earlier \emph{From Search to Research} studies model capability and research budgets in a controlled benchmark \citep{deng2026search}; the present paper focuses on the framework and its continuing production process.

The production records show two ways in which research continues after initialization. The agent explores new directions and revisits candidates from earlier investigations, including report-grounded proposals that were not initially admitted. Later admissions broaden the library toward fundamental and market-interaction constructions. Tests and screening determine which proposals survive: a proposed refinement can lower its parent's score, and a high-scoring candidate can be rejected as redundant.

We evaluate the resulting pooled library of 105 admitted factors. Its median 1-day Rank IC is 0.0301. A real case links an AI-generated idea to six implemented versions, and a weekly portfolio of the combined library signals returns 16.81\% annualized after stated costs over 2020--2026, compared with 1.71\% for CSI 1000. The library and portfolio are retrospective illustrations, rather than discovery-time out-of-sample tests. The production records, case study, and portfolio connect research decisions to individual factors and their combined use.

This paper contributes an initial framework, an account of its research process, and evidence about its resulting factors. The next questions concern how initialization changes the idea search space, whether experience improves subsequent research, and how the yield of new discoveries changes as search effort and the existing library grow. Controlled comparisons belong to future benchmarks. Extending the loop from factor construction to autonomous strategy research is a further direction.

\begin{figure}[t]
\centering
\includegraphics[width=\textwidth]{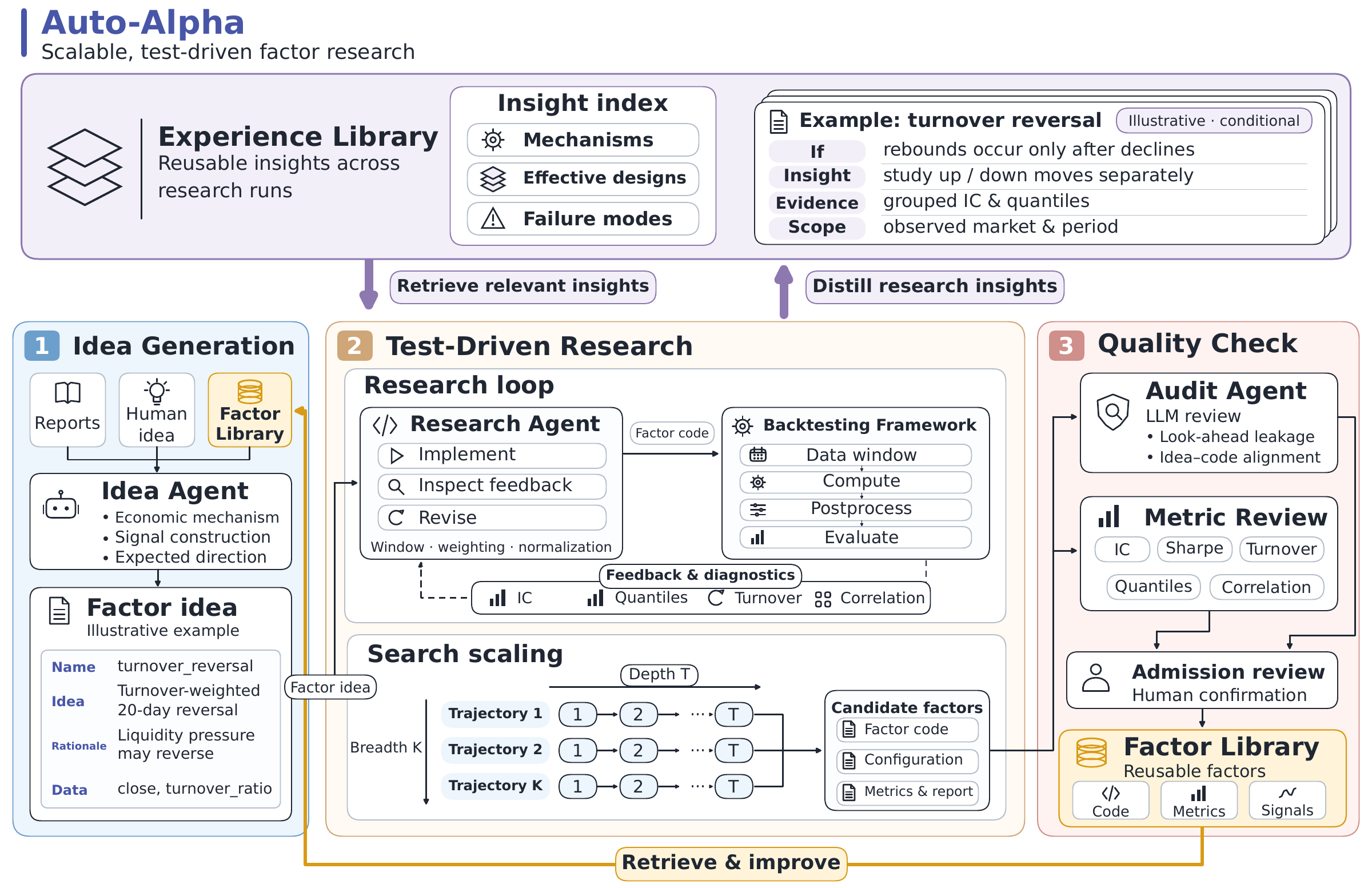}
\caption{\textbf{Auto-Alpha's factor-research architecture.} Report-grounded proposals bootstrap the workflow; subsequent ideas can draw on the Factor Library and Experience Library. Backtesting sits inside the research loop, where the agent implements, inspects, and revises. Quality review combines LLM audit, quantitative checks, and human admission. The two-stage production view appears in Figure~\ref{fig:two-stage-funnels}. The turnover-reversal cards illustrate a proposed idea and a conditional lesson; they are not additional measured findings.}
\label{fig:framework}
\end{figure}

%% file: sections/related_work.tex
\section{Related Work}
\paragraph{From formula search to agent-led factor research.}
Automated factor discovery predates language-model agents. AutoAlpha uses hierarchical evolutionary search to construct factors \citep{zhang2020autoalpha}; despite the similar name, it is a separate system from the Auto-Alpha workflow described here. AlphaGen optimizes a collection of formulaic factors through reinforcement learning, using their contribution to a downstream combination model as the objective \citep{yu2023alphagen}. AlphaForge couples factor generation with dynamic combination \citep{shi2024alphaforge}. These approaches establish search and factor complementarity as central concerns. Auto-Alpha uses a general-purpose code interface and records the surrounding research trajectory: how an agent proposes a mechanism, tests an implementation, and decides what to revise.

Language-model systems extend factor discovery toward interactive and iterative research. Alpha-GPT 2.0 incorporates human feedback into alpha development \citep{yuan2024alphagpt}. CogAlpha evolves code-based factors through refinement, mutation, and recombination \citep{liu2026cogalpha}. R\&D-Agent-Quant separates research and development and studies joint factor and model optimization \citep{li2025rdagent}, while AlphaAgent regularizes originality and complexity during exploration \citep{tang2025alphaagent}. Auto-Alpha shares the use of model reasoning, executable artifacts, and evaluation feedback. Its scope is factor mining within a supplied environment, with within-trajectory revisions delegated to the agent rather than organized around a prescribed catalogue of evolutionary operators.

\paragraph{Grounding and accumulated research experience.}
Persistent knowledge is already a substantive component of agentic factor research. X-Alpha abstracts quantitative reports into memory, uses memory to guide subsequent research, and consolidates feedback across cycles \citep{liu2026xalpha}. FactorMiner combines modular skills with structured experience in a retrieve--generate--evaluate--distill loop, including successful patterns and failure constraints \citep{wang2026factorminer}. Both are directly relevant to our Experience Library; report grounding and cross-run knowledge reuse are therefore shared design elements. We include these mechanisms in an operational framework and examine their use alongside production accounting and factor outputs. The available records establish instances of retrieval and application, but do not isolate an independent performance gain from experience accumulation.

\paragraph{Tool feedback and autonomous research.}
The research loop belongs to a broader family of agents that use observations to choose subsequent actions. ReAct interleaves reasoning with environment interaction \citep{yao2023react}, and Reflexion retains verbal feedback to inform later attempts without updating model weights \citep{shinn2023reflexion}. The AI Scientist extends automation across idea generation, implementation, experimentation, and paper preparation \citep{lu2024aiscientist}. Auto-Alpha focuses on a narrower artifact and environment: a factor implementation evaluated through supplied data and backtests. We use \emph{test-driven research} to describe the agent's dependence on executable diagnostics, including evidence against its hypothesis. Neither successful execution nor a favorable historical metric, by itself, establishes a valid economic explanation.

\paragraph{Search scaling and budget allocation.}
Inference-time scaling distinguishes generating more candidates from selecting a better result. Repeated sampling expands candidate coverage, but its practical benefit depends on effective selection \citep{brown2024monkeys}. Compute-aware inference studies examine the allocation of effort to search and refinement \citep{snell2025scaling}; agent-focused work studies parallel sampling and sequential revision \citep{zhu2025agentscaling}. These distinctions motivate our separation of research breadth and depth, without implying that their gains transfer unchanged to noisy financial feedback. Our earlier study, \emph{From Search to Research}, evaluates models on 50 report-grounded factor tasks and examines search budgets \citep{deng2026search}. The present paper addresses the surrounding workflow that proposes tasks, accumulates experience, and reviews candidates. Its retrospective records do not extend the earlier benchmark's quantitative conclusions.

\paragraph{Evaluating research capability.}
Research benchmarks differ in the decisions they leave to the agent. PaperBench evaluates replication of published research \citep{starace2025paperbench}; RE-Bench evaluates research engineering under resource budgets \citep{wijk2025rebench}. ResearchGym asks agents to develop methods within supplied research environments \citep{garikaparthi2026researchgym}, while PostTrainBench studies agent-directed model post-training \citep{rank2026posttrainbench}. These settings motivate evaluating a sequence of research decisions, while keeping the artifact, environment, and budget explicit.

Financial benchmarks already address complementary parts of this problem. AlphaBench evaluates formulaic factor generation, evaluation, and search \citep{luo2026alphabench}. AlphaQT-Bench separates executable code from quantitative correctness through checks of causality, functionality, and structural constraints \citep{luo2026alphaqt}. OpenFinGym provides verifiable environments across financial tasks \citep{zhang2026openfingym}. Our proposed evaluation agenda builds on this distinction between artifact performance and research reliability, adding matched initialization sources and versioned experience and factor libraries. It also inherits the statistical difficulty of selecting candidates after repeated backtests \citep{bailey2017overfitting}; an LLM audit does not remove that difficulty. This paper specifies those questions rather than introducing another evaluated benchmark.

%% file: sections/setting.tex
\section{Research Setting}
\label{sec:setting}
In this paper, a factor is an executable scoring rule that uses information available at a given time to rank stocks by their expected relative returns over a specified horizon. Applied to the eligible stock universe, it produces one score per stock with sufficient data, inducing a cross-sectional ordering. After aligning the signal direction, higher scores express a hypothesis of stronger subsequent returns relative to lower-scoring stocks. The economic rationale explains why this ordering might be informative; its predictive value must be tested. Section~\ref{sec:empirical} evaluates this ordering with Rank IC and uses combined factor scores to select a Top-300 portfolio.

The full factor-mining task starts with a supplied environment and includes generating the ideas to investigate. The idea agent proposes a factor idea and its economic rationale using domain reports, human research directions, or existing factors as source material. That proposal then initiates a research trajectory of implementation, testing, and revision. The environment contains a data catalogue, access conventions, factor execution interfaces, postprocessing options, and a backtesting framework. The research agent may change the factor implementation and permitted configuration, including selected windows, weighting, and postprocessing choices. The evaluator is supplied infrastructure rather than an object for the agent to redesign. 

\paragraph{Data setup.}
The supplied environment covers mainland Chinese A-share equities. It provides daily prices, volume, turnover, valuation and capitalization fields; intraday bars at 1, 5, 15, 30, and 60 minutes; daily order-size money flows; quarterly financial statements; and announcement-driven earnings forecasts and preliminary results. Industry histories, daily style exposures, and index data provide additional context and controls. Factors convert these inputs into daily cross-sectional signals using information available by the end of the signal day. Financial records are exposed by disclosure date, not simply by their accounting period. The default stock universe excludes Beijing-listed stocks, special-treatment (ST) stocks, and stocks listed for fewer than 60 calendar days. The library diagnostics span 2013--2026, while the portfolio illustration starts in 2020; individual fields and stocks have different coverage. Appendix~\ref{sec:data-setup} details the sources, frequencies, timing rules, and study snapshots.

A trajectory is a sequence of implementations and tests associated with one research idea. At iteration $t$, the agent proposes an implementation $f_t$ and an allowed configuration $c_t$, receives diagnostics $y_t=\mathcal{B}(f_t,c_t;\mathcal{D})$, and decides how to proceed. Here $\mathcal{D}$ denotes the supplied data and $\mathcal{B}$ the evaluation framework. Diagnostics include information coefficients (IC), portfolio performance, quantile behavior, turnover, and similarity to existing factors when available. These signals address different questions: predictive association, portfolio behavior, implementation costs, and potential redundancy.

Research success combines empirical promise with a defensible construction. A high score alone does not establish that an implementation measures the proposed mechanism, uses only information available at the relevant time, or adds useful diversity to the library. The agent therefore delivers an inspectable artifact containing code, configuration, metrics, and a research report. A failed or unsupported hypothesis can still produce a useful research record, even when its factor is not suitable for admission.

We examine three operational questions: how report-guided and self-directed research produce inspectable artifacts, which outputs survive performance and novelty review, and how the admitted factors behave individually and in combination. Production records trace proposal development and admission, while common-snapshot diagnostics evaluate the resulting library.

%% file: sections/approach.tex
\section{The Auto-Alpha Framework}
\label{sec:approach}
Auto-Alpha separates the research decisions made by agents from the data and evaluation infrastructure supplied by the environment. Its persistent libraries connect otherwise separate investigations.
\subsection{From Report-Guided Bootstrapping to Self-Directed Research}
The first stage turns quantitative reports into starting points for research. The idea agent identifies an economic mechanism, maps it to available observables, and proposes a construction to test. The research agent can adapt the construction as it implements and evaluates it; the task is not limited to copying a report's formula. Reviewed artifacts enter the Factor Library, and research records yield conditional insights for the Experience Library. A rejected candidate may still contribute a useful diagnostic lesson.

The second stage shifts the immediate source of proposals to the idea agent's exploration of the accumulated libraries. Existing factors can expose combinations or mechanisms worth investigating, while experience entries can suggest applicable constructions or pitfalls. Each new idea enters the same research and assessment process. Further factors and lessons update the resources available to later attempts. 

\subsection{Generating Economically Motivated Ideas}
The idea agent proposes an economic mechanism and a concrete signal construction to investigate. Quantitative research reports provide economic explanations, observable proxies, and examples of signal construction. Human directions can specify a question or mechanism, while an existing factor can anchor an improvement task. A generation request selects one primary source type; library inspection can provide additional context for any of them. The idea agent also receives the data catalogue as context. The current prompt permits proposals that require unavailable fields, leaving feasibility and possible proxies to the research stage.

A factor idea connects an economic explanation to an observable construction. The generation prompt asks for an idea and rationale, the required data, and source information. A useful rationale explains why the proposed signal might predict subsequent returns, which direction is expected, and what would make that interpretation questionable. This requirement discourages arbitrary combinations of variables without a research question. It is still a prompting constraint: feasibility and semantic consistency must be checked during implementation and review.

For example, the saved \texttt{neglect\_ato\_improver} idea combines improving quarterly asset turnover with low market attention: operating-efficiency improvements might be priced in more slowly for neglected stocks. It proposes a product of the two cross-sectional ranks and identifies revenue, assets, turnover, and capitalization as inputs. Appendix~\ref{sec:case-idea} gives the structured idea; the subsequent subsections trace its implementation, feedback, delivery, and audit.

\subsection{A Simple Harness for Test-Driven Research}
The harness supplies the research environment while the agent controls the research decisions. It exposes data conventions, execution commands, diagnostic outputs, and artifact locations. The research agent is responsible for forming a concrete implementation, invoking backtests, inspecting results, and choosing the next action. 

Each iteration connects a proposed change to an observable diagnostic. The agent may first repair an invalid computation or timing error, then test a construction choice such as a normalization, window, or conditional filter. After execution, it reads feedback and records what changed, what was observed, and why another revision is warranted. A useful step narrows an uncertainty: for example, whether a sparse event ratio measures the intended conditional effect or mainly proxies event frequency. The backtesting framework belongs inside this loop because its feedback determines subsequent research actions. This follows the general reasoning--action--observation structure of tool-using agents \citep{yao2023react}, with financial diagnostics supplying the observations.

Test-driven research requires interpreting results rather than optimizing one displayed metric blindly. Changes in IC, quantile ordering, turnover, and exposure can contradict an attractive headline Sharpe ratio. The agent can preserve a stronger earlier implementation or document that the original hypothesis remains unsupported. These are empirical research tests rather than software unit tests, and repeated use of the same evaluation period can induce selection bias \citep{bailey2017overfitting}. Appendix~\ref{sec:case-feedback} shows the actual feedback and the next research decisions in one trajectory. Separate assessment is therefore necessary even when every iteration executes correctly.

\subsection{Test-Driven Search Scaling}
Search effort can grow along the depth of a trajectory and the breadth of initiated trajectories. Depth provides additional opportunities to investigate one hypothesis; breadth introduces alternative starting points or independent attempts. If $K$ trajectories each receive up to $T$ iterations, $KT$ describes a nominal iteration budget, but actual compute also depends on data operations, execution failures, model calls, and diagnostic work. Equal iteration counts therefore need not imply equal costs. This distinction is consistent with work on compute allocation and parallel versus sequential agent search \citep{snell2025scaling,zhu2025agentscaling}.

Auto-Alpha leaves within-trajectory revisions to the agent and uses run orchestration to launch additional research. Iteration limits guide the research session, and the agent selects a candidate based on its reasoning and feedback. The implementation's prompt-level budget is not a strict count of all backtest invocations. Table~\ref{tab:case-versions} illustrates depth through six saved versions; Section~\ref{sec:throughput} describes continued proposal generation and its outputs.

\subsection{The Experience Library}
The Experience Library stores reusable research knowledge beyond any individual run. A run log records what happened in a particular investigation; an experience entry attempts to explain when a construction or diagnostic is useful and what evidence supports that lesson. Relevant content includes mechanism interpretations, effective designs, and failure modes. The current realization uses a searchable Markdown index and entries that the agent can inspect with ordinary tools.

The desired unit of reuse is a conditional insight with a traceable origin. For example, a finding that a sparse event ratio is dominated by event frequency should motivate separate tests of the event count and the conditional magnitude before further ratio tuning. An entry should retain the setting in which the observation arose and the source run. During a later investigation, the agent can retrieve the insight, assess whether its assumptions apply, and test an associated action. Research can then add or revise an entry, making knowledge accumulation a cross-run process. The principle is related to verbal feedback reuse in Reflexion and structured financial experience in FactorMiner \citep{shinn2023reflexion,wang2026factorminer}; our emphasis is retaining the conditions and evidence behind each lesson. Appendix~\ref{sec:experience-examples} gives actual retrieval, selective-use, and writeback examples.

Experience remains provisional until its scope and consequences are checked. A lesson can be inapplicable to another mechanism, reflect an earlier evaluation artifact, or encode an overgeneralization from one run. Source records and conditions of applicability are important context when evaluating a retrieved lesson.

\subsection{Quality Check and Factor Admission}
Candidate assessment combines semantic audit with quantitative review (see the worked assessment in Appendix~\ref{sec:case-assessment}). A separate LLM audit examines issues such as look-ahead leakage and alignment between the idea and its implementation. Metric review considers performance and diagnostic behavior, including IC, Sharpe ratio, turnover, quantiles, and correlation with existing factors. These checks provide complementary evidence: economic plausibility cannot establish empirical quality, while a strong backtest cannot establish correct information timing or hypothesis fidelity.

Admission to the Factor Library remains a human-confirmed decision. The current system saves a candidate before audit and appends audit findings to the research record; it does not implement the entire quality check as an unbypassable automatic gate. The library contains reusable factor artifacts, while the Experience Library retains lessons that may also arise from rejected candidates. Figure~\ref{fig:framework} summarizes the review workflow and resulting information flows.

%% file: sections/empirical.tex
\section{Two-Stage Production and Factor Quality}
\label{sec:empirical}
\input{sections/throughput}

\subsection{Quality of the Discovered Factors}
We evaluate a fixed library snapshot of 105 admitted factors: 69 technical, 35 fundamental, and one classified as other. The latest mining batch contributes to the production counts above but is not included in this factor-quality and portfolio snapshot. The snapshot evaluations cover 2013-01-04 through 2026-11-03. Table~\ref{tab:library-ic} summarizes the full library, with individual examples in Appendix~\ref{sec:factor-examples}. Across the library, the median 1-day Rank IC is 0.0301 and the maximum is 0.0651; the median unannualized 1-day ICIR is 0.530. These diagnostics characterize the selected library's historical predictive content.

\input{tables/library_summary}

Stage-specific composition shows how self-directed research extends the cold start. The 32 report-guided admissions comprise 30 technical, one fundamental, and one other factor. The 43 self-directed admissions in the evaluation snapshot comprise 17 technical and 26 fundamental factors. These broad metadata labels describe primary information content and include fundamental--market interactions. Appendix~\ref{sec:phase-quality} shows the stage distributions under the same evaluation snapshot. Since proposal directions, models, and budgets differ, the composition change is descriptive rather than a ranking of stage-level research ability.

We use 1-day Rank IC as the primary predictive metric, matching the daily return horizon used for research optimization and rolling combination weights. For a signal computed at the close of day $t$, the $h$-day evaluation label is $P^{\mathrm{VWAP}}_{t+1+h}/P^{\mathrm{VWAP}}_{t+1}-1$. The primary label sets $h=1$, measuring the return from day $t+1$ VWAP to day $t+2$ VWAP. Rank IC is the cross-sectional Spearman correlation on stocks with both signal and label available. We retain each factor's admitted postprocessing configuration: most use industry and size neutralization followed by ranking, while some include additional controls. Thus, the table describes the library as used, rather than a comparison under one newly imposed neutralization specification. Supplementary 5- and 10-day diagnostics are reported in Appendix~\ref{sec:supplementary-horizons}. The one-day evaluation horizon is distinct from factor lookback windows and the portfolio's weekly rebalancing schedule.

\input{tables/factor_type_sentence}

\begin{center}
\begin{minipage}{\textwidth}
\captionsetup{hypcap=false}
\centering
\includegraphics[width=\textwidth]{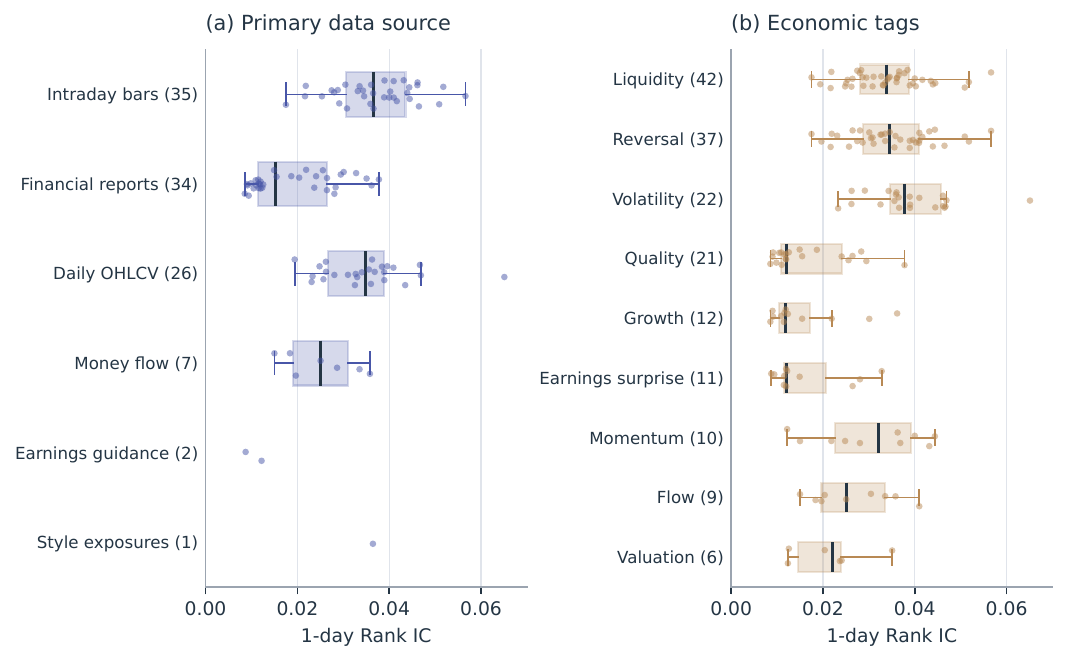}
\captionof{figure}{\textbf{Distribution of 1-day Rank IC across factor types.} Each point is one admitted factor; labels give group counts. Left: the saved primary-data labels partition the full library. Right: existing economic tags with at least five factors; a factor may appear under multiple tags. Boxes show the interquartile range and median, with whiskers extending to observations within 1.5 interquartile ranges. Groups with fewer than five factors are shown as points only. All values use the same evaluation period as Table~\ref{tab:library-ic}.}
\label{fig:factor-types}
\end{minipage}
\end{center}

\subsection{Selecting Factors from Research Outputs}
\label{sec:library-increment}
\begin{figure}[htbp]
\centering
\includegraphics[width=\textwidth]{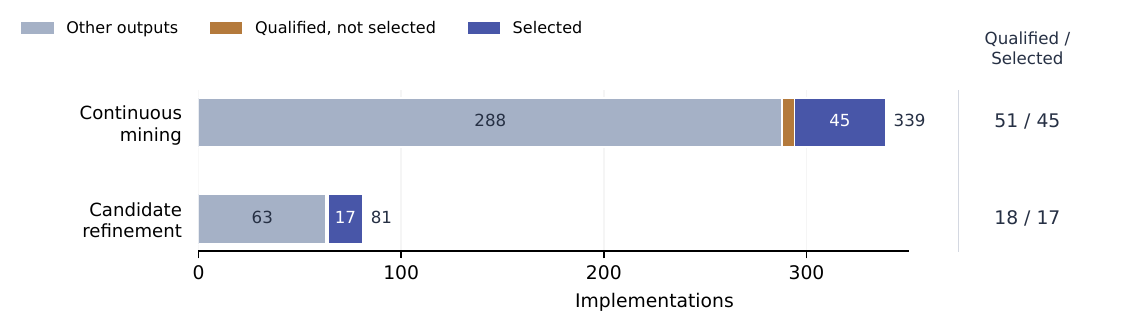}
\caption{\textbf{From implementations to selected factors.} Bars partition outputs into other outputs, qualified candidates not selected, and selected factors. Numbers at right give qualified/selected counts; selection follows the definition in Table~\ref{tab:production-funnels}.}
\label{fig:route-outcomes}
\end{figure}

Performance screening removes most recorded implementations before library admission. Across 420 self-directed implementations, 69 meet the Sharpe threshold and 62 are selected. Figure~\ref{fig:route-outcomes} shows the outcomes for continuous mining and candidate refinement.

Similarity review distinguishes strong standalone scores from useful library additions. In one new-direction screening batch, 18 candidates qualify but four are rejected: two overlap with existing factors and two with other candidates selected in the same batch. The remaining 14 enter the library. Candidate refinement provides another source of additions by developing earlier research into new constructions; Section~\ref{sec:insights} examines these inherited proposals and their outcomes.

\subsection{A Traceable Idea-to-Factor Example}
The factor \texttt{neglect\_ato\_improver\_v6} provides a concrete example of the full workflow. Its saved initial idea combines improving quarterly asset turnover with low market attention, reasoning that improvements in operating efficiency may be incorporated into prices slowly for neglected stocks. The idea agent proposed a rank product of these two components after inspecting existing factors for underexplored combinations. The research trajectory retained this mechanism while testing data freshness, the turnover window, and the use of earlier earnings disclosures. The delivered implementation uses a 10-day attention window, compared with approximately 60 days in the original proposal. Appendix~\ref{sec:real-case} gives the actual idea, the factor construction, the six archived versions, and the audit's recorded deviations. In the current library snapshot, this factor has 1-day Rank IC 0.0378 and unannualized 1-day ICIR 0.538. Figure~\ref{fig:real-factor-diagnostics} in Appendix~\ref{sec:factor-diagnostics} shows its long--short and decile diagnostics.

\subsection{A Simple Portfolio of the Discovered Factors}
We combine all factors in the library with a rolling maximum-ICIR rule and hold an equal-weight portfolio of the 300 highest-scoring eligible stocks. Factor weights are nonnegative and estimated from at most 252 past daily IC observations, with a minimum of 126 observations and 10\% diagonal covariance shrinkage. Writing $\mu$ for the mean daily IC vector and $\widehat\Sigma$ for its shrunk covariance, the IC-space objective is
\begin{equation}
 \max_{w\geq 0,\;\mathbf{1}^{\top}w=1}
 \frac{w^{\top}\mu}{\sqrt{w^{\top}\widehat\Sigma w}}.
\end{equation}
We refit every 20 trading days using the repository's existing maximum-ICIR solver. At close $t$, training includes only labels associated with signals dated $t-2$ or earlier, whose one-day forward VWAP returns have already matured. Direction-adjusted factor values are transformed into centered percentile ranks before combination; missing signals contribute zero. The same fixed library of approximately 100 factors is used throughout, and the optimizer may assign zero weight to a factor. Figure~\ref{fig:maxicir-diagnostics} in Appendix~\ref{sec:factor-diagnostics} reports the combined signal's daily long--short and decile diagnostics.

The account runs from the first trading day of 2020 through 2026-11-03. Each week's first trading day produces a new Top-300 target for execution on the next trading day. The eligible Shanghai and Shenzhen A-share universe excludes ST stocks and stocks listed for fewer than 60 days; actual fills respect the engine's suspension, price-limit, and approximate lot-size rules. No additional turnover or stock-replacement limit is imposed. Pending orders may be retried between rebalances, and ST exits are handled separately. Initial capital is RMB 100 million. Constant fees throughout the period are 1 basis point for buys and 6 for sells, with a RMB 5 minimum per stock and direction; proportional slippage adds 10 basis points on each side. Trades use VWAP and the account is marked at the close. This execution model does not estimate market impact or impose a volume-participation cap. CSI 1000 (000852.SH) is the reference benchmark, with no index hedge or constraint restricting selection to its constituents. Eight missing benchmark sessions in the local index file were recovered from the same Tushare source; all seven overlapping sessions in the recovery query matched the stored closes.

\input{tables/portfolio_summary_sentence}

\input{tables/portfolio_results}
The execution ledger retains 2 positions without a fresh quote for more than 60 trading sessions, valued at their last observed prices. Together they represent 0.53\% of terminal account value. Writing off those terminal values reduces the annualized return to 16.71\%. This is a terminal-valuation sensitivity, not a replay using historical delisting dates.

This exercise demonstrates how the library's signals translate into an executable portfolio under a simple construction. It is a retrospective historical illustration: the current factor definitions, directions, and admission decisions were developed using historical data, including periods shown here. Rolling coefficient estimation prevents future IC observations from entering a particular weight fit; it does not make the preceding factor discovery and selection out of sample. In addition, stock returns reflect dividends through adjusted prices, whereas the CSI 1000 reference is a price index, and the portfolio has no constraints matching the index's size or industry exposures. The comparison therefore reports realized backtest behavior under this protocol, rather than isolating stock-selection alpha or a causal benefit of the agent architecture.

\begin{center}
\begin{minipage}{\textwidth}
\captionsetup{hypcap=false}
\centering
\includegraphics[width=\textwidth]{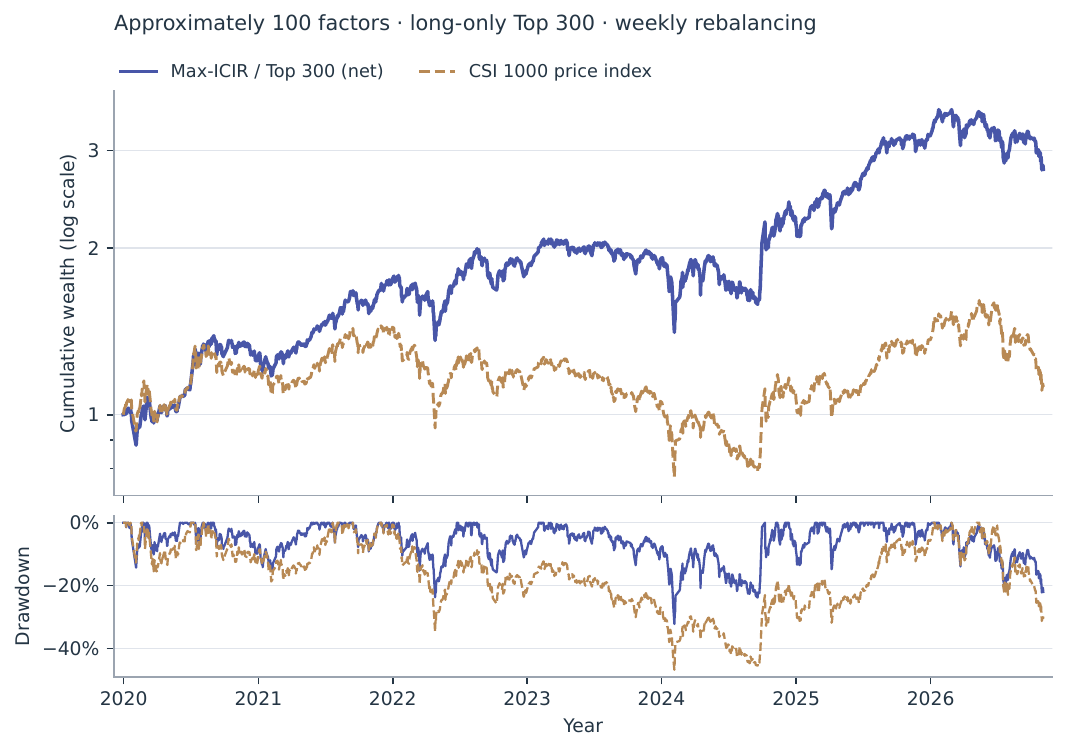}
\captionof{figure}{\textbf{Weekly Top-300 portfolio using the full discovered factor library.} The strategy includes the stated fees and slippage; the benchmark is the CSI 1000 price index. Top: cumulative wealth on a logarithmic scale. Bottom: drawdowns from prior peaks. Factor-combination weights use only matured historical labels, but the factor library itself was selected retrospectively.}
\label{fig:portfolio}
\end{minipage}
\end{center}

%% file: sections/throughput.tex
\subsection{From Report-Guided Ideas to Continued Research}
\label{sec:throughput}
\begin{figure}[htbp]
\centering
\includegraphics[width=\textwidth]{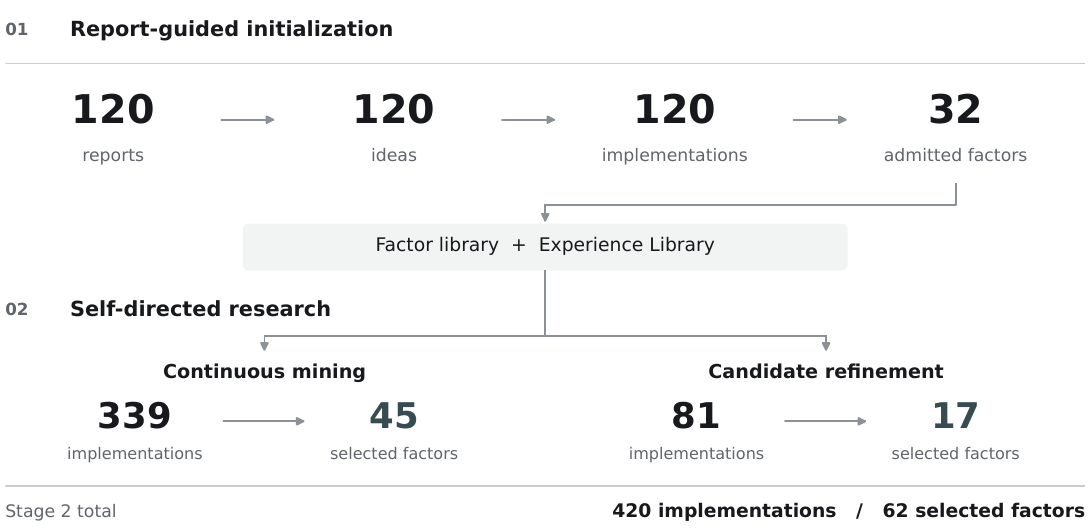}
\caption{\textbf{From report-guided initialization to continued factor research.} The idea agent continuously generates new proposals and revisits existing candidates. Report-derived parents seed nine later selections. Counts summarize recorded implementations and selection outcomes.}
\label{fig:two-stage-funnels}
\end{figure}

We implement 120 factor ideas drawn from 120 quantitative research reports, yielding 32 admitted factors. Subsequent research follows two routes: continuous mining of new ideas and refinement of existing candidates (Figure~\ref{fig:two-stage-funnels}). Both use the same implementation, testing, and review process.

Continuous mining produces 339 implementations, of which 51 meet the Sharpe threshold and 45 are selected after screening. Candidate refinement produces 81 implementations, 18 qualified candidates, and 17 selections. Together, self-directed research yields 420 implementations and 62 selected factors (Table~\ref{tab:production-funnels}). Appendix~\ref{sec:production-accounting} describes the underlying records and the latest batch summary.

\input{tables/production_funnels}

%% file: tables/production_funnels.tex
\begin{table}[htbp]
\centering\small
\caption{\textbf{Research outputs and selection outcomes.} Qualification uses a net long--short Sharpe threshold of 3.}
\label{tab:production-funnels}
\begin{tabularx}{\textwidth}{@{}Xrrr@{}}
\toprule
Research activity & Outputs & Qualified & Selected \\
\midrule
Report-guided initialization & 120 & 37 & 32\textsuperscript{a} \\
\midrule
Self-directed: continuous mining & 339 & 51 & 45 \\
Self-directed: candidate refinement & 81 & 18 & 17 \\
\textbf{Self-directed total} & \textbf{420} & \textbf{69} & \textbf{62}\textsuperscript{b} \\
\bottomrule
\end{tabularx}
\vspace{3pt}
\parbox{\textwidth}{\footnotesize a: Retention is counted by idea lineage. b: Selections comprise 43 recorded admissions and 19 candidates from the latest batch passing both Sharpe and correlation (below 0.7) screens.}
\end{table}

%% file: tables/library_summary.tex
\begin{table}[htbp]
\centering\small
\caption{\textbf{Predictive diagnostics across the pooled library of 100+ admitted factors.} The primary horizon is one trading day. Snapshot evaluated over 2013-01-04--2026-11-03. IC is daily cross-sectional Spearman correlation with the indicated forward VWAP return; ICIR is its time-series mean divided by standard deviation, without annualization. Quantiles are across factors. These are retrospectively selected library artifacts.}
\label{tab:library-ic}
\begin{tabular}{@{}lrrrrr@{}}
\toprule
Horizon & IC 25th pct. & Median IC & IC 75th pct. & Maximum IC & Median ICIR \\
\midrule
1 day & 0.0204 & 0.0301 & 0.0378 & 0.0651 & 0.530 \\
\bottomrule
\end{tabular}
\end{table}

%% file: tables/factor_type_sentence.tex
The library spans several data sources and economic construction types. Using its existing broad labels, the median 1-day Rank IC is 0.0356 for 69 technical factors and 0.0126 for 35 fundamental factors. Figure~\ref{fig:factor-types} shows the individual-factor distributions by primary data source and by economic tag. Data-source groups partition the library; economic tags overlap because one construction can combine, for example, quality and liquidity. These are descriptions of admitted artifacts. Without type-specific counts of attempted and rejected ideas, group differences do not measure discovery difficulty or admission yield.

%% file: tables/portfolio_summary_sentence.tex
The net portfolio returns 16.81\% annualized, compared with 1.71\% for the CSI 1000 price index. Its Sharpe ratio is 0.763, with a maximum drawdown of 32.16\%. Annualized one-way turnover is 19.15 times account value. The combined score has 1-day Rank IC 0.0471 and unannualized ICIR 0.947 over the portfolio evaluation period. Table~\ref{tab:portfolio} reports the account metrics and Figure~\ref{fig:portfolio} shows the wealth and drawdown paths.

%% file: tables/portfolio_results.tex
\begin{table}[htbp]
\centering\small
\caption{\textbf{Net portfolio performance versus CSI 1000, 2020-01-02--2026-11-03.} Return and volatility are annualized with 252 trading days; Sharpe uses zero risk-free return. Strategy results include fees and slippage. Turnover is one-sided traded value relative to account value, annualized. The index is a price-return reference without transaction costs.}
\label{tab:portfolio}
\begin{tabular}{@{}lrr@{}}
\toprule
Metric & Max-ICIR Top 300 & CSI 1000 \\
\midrule
Annualized return & 16.81\% & 1.71\% \\
Annualized volatility & 24.26\% & 25.12\% \\
Sharpe ratio & 0.763 & 0.193 \\
Maximum drawdown & -32.16\% & -46.71\% \\
Annualized turnover (times) & 19.15 & --- \\
Information ratio vs. CSI 1000 & 1.552 & --- \\
\bottomrule
\end{tabular}
\end{table}

%% file: sections/insights.tex
\section{Findings from Continued Factor Research}
\label{sec:insights}
Production lineages and research trajectories reveal how initial candidates support later proposals, how tests revise economic constructions, and what knowledge persists after an investigation. Appendix~\ref{sec:provenance} identifies the case material and its interpretation.

\subsection{Cold-Start Candidates Become Seeds for Further Research}
Report guidance supplies concrete choices for an agent to examine: a mechanism, an observable proxy, a comparison group, and a relevant time scale. The agent must still translate these choices into available fields and executable operations. We use \emph{search-space initialization} to describe this influence on proposed starting points, rather than a hard restriction on the programs the research agent may write. The two-stage workflow makes that initialization operational; the subsequent proposals can extend or revise the initial constructions.

The connection between stages is visible in subsequent refinement. Of 78 distinct parent candidates selected for a refinement campaign, 45 are final deliveries from the report-guided stage. New proposals based on those parents lead to nine admitted child factors. For example, research on daily price--volume correlation leads to a later price-location and volume-response construction, while an abnormal-attention reversal candidate seeds an asymmetric attention construction. These descendants show how initial candidates remain useful research material even when they do not enter the library in their original form.

\subsection{Ideas Combine Economic Mechanisms with Testable Constructions}
The ideas reveal recurring research operations that make the initialization hypothesis more concrete. Table~\ref{tab:patterns} organizes examples by the transformation applied to a mechanism, rather than by a financial topic label. A proposal often becomes more specific by asking when a familiar mechanism should apply, which independent evidence could validate it, or which component has already been reflected in prices. These patterns are overlapping qualitative categories, and their stated mechanisms remain hypotheses.

A recurring construction in the later production episode conditions fundamental changes on market response. Ten of its 14 admitted factors combine a fundamental signal with an attention, price, or flow condition. Examples include operating-efficiency improvement among neglected stocks, an earnings surprise accompanied by accumulation, and revenue improvement that has not been followed by a price increase. These factors extend the proposal beyond a standalone accounting ratio by specifying when the economic information may remain incompletely reflected in prices.

\begin{table}[t]
\centering
\small
\caption{Illustrative idea patterns identified in existing records. The descriptions summarize proposed research mechanisms, not validated economic laws or a frequency ranking.}
\label{tab:patterns}
\begin{tabularx}{\textwidth}{@{}>{\raggedright\arraybackslash}p{.23\textwidth}X@{}}
\toprule
Research operation & Example of a testable question \\
\midrule
Condition a mechanism & On price-shock days, does large-versus-small order flow differ meaningfully from its normal-day baseline? \\
Seek independent confirmation & Does labor investment help distinguish the quality of an earnings surprise? \\
Remove an already reflected component & Is an earnings revision informative after accounting for associated price adjustment? \\
Separate asymmetric states & Do similar flow or volatility measures have different meanings following upward and downward moves? \\
Choose a mechanism-based clock & Does reversal become clearer when its window is defined by cumulative turnover? \\
Change the reference set & Does comparison with a firm's own normalized margins or peer disclosure breadth clarify the signal? \\
\bottomrule
\end{tabularx}
\end{table}

Grounding is useful as a design principle because it can supply a structure that the model can continue to manipulate. The chain is economic mechanism, observable proxy, meaningful comparison, and executable test. Producing a fluent financial explanation is insufficient if the model cannot align those parts or recognize when the evidence contradicts the explanation. Initialization and subsequent research capability are therefore complementary objects of study.

\subsection{Tests Support Revision and Rejection, Not Guaranteed Improvement}
Candidate-guided research often fails to improve the parent's score. In 73 recorded parent--child comparisons, 30 child scores increase and 43 decrease; the median net Sharpe difference is $-0.139$ (Appendix~\ref{sec:refinement-outcomes}).  The 17 admissions from this route emerge through selection among these proposed revisions.

The selected trajectories include both continued improvement and early plateaus. In an asset-turnover improvement run, the main sequence of six named candidate versions rises from a recorded net Sharpe of 3.189 to 3.605, with smaller gains in the final two steps. In a gross-profit/earnings-surprise run, the second of eight recorded versions achieves 3.345 after an initial 2.954; subsequent variants range from 3.009 to 3.325 and do not exceed it. The agent retains the earlier version. These cases illustrate how additional depth can refine a useful idea or expose that further local changes are unproductive.

A failed hypothesis illustrates a different form of useful research behavior. In a loss-recognition asymmetry run, the versions consistent with the stated hypothesis remain negative in the recorded tests. The agent documents this failure instead of automatically reversing the signal to pursue a positive score. Such a trajectory is not a successful factor discovery, but it preserves the distinction between testing the proposed explanation and changing the question after seeing the result.

Our earlier benchmark studies model and budget differences systematically \citep{deng2026search}; the trajectories here illustrate the decisions made within individual investigations.

Tests also expose implementation assumptions that economic reasoning alone does not resolve. In the flow-impact asymmetry investigation, the agent diagnoses an almost-zero explanatory variable: subtracting aggregate buy-side and sell-side amounts in the supplied table largely cancels the same traded volume. It replaces that construction with a nondegenerate net-flow field, but the resulting factor remains below the admission threshold. Its research card recommends inspecting the scale and variation of a new predictor before spending iterations on downstream normalization. 

\subsection{Experience Informs Later Research Decisions}
Experience reuse is visible in concrete research decisions. The examples in Appendix~\ref{sec:experience-examples} connect retrieved entries to proposed changes, evaluation feedback, and selective rejection. 

A more informative trace links retrieved experience to a subsequent research action. In the gross-profit/earnings-surprise run, the agent reads an entry about postprocessing expansion-confirmation factors, cites it when planning a change, and records an improvement from 2.954 to 3.345 in the next version. A separate asset-turnover run considers the scope of related advice and finds that some corresponding changes do not help. Together, these records illustrate retrieval, interpretation, testing, and selective rejection. 

Research also generates reusable diagnostic procedures. A limit-resilience run investigates whether a sparse event ratio is driven by conditional magnitude or by event frequency, then writes an experience entry directing future agents to separate these components first. This example illustrates the intended transition from a run-specific result to a research insight. 

The loss-recognition record also illustrates a knowledge-quality concern: its agent-written interpretation generalizes a single trajectory into a broad market explanation. Preserving hypothesis fidelity in candidate selection does not validate every explanation in the research notes. An experience entry needs to retain this distinction, so that a later agent can test the proposed explanation with its evidential scope in view.

%% file: sections/discussion.tex
\section{Boundaries and Future Research}
\label{sec:discussion}
Auto-Alpha provides an operating interface between economic ideas, executable factor research, and accumulated artifacts. Its two-stage workflow is a practical organization of research, not a controlled comparison between initialized and uninitialized agents. Human admission remains part of the process, LLM audits can miss errors, and an economically plausible explanation need not be empirically correct. Separating execution success from quantitative correctness remains essential \citep{luo2026alphaqt}.

\paragraph{Measuring the idea search space.}
A future benchmark can distinguish unguided idea generation, direct report-guided generation, and autonomous generation after report-based bootstrapping. It should measure idea validity, mechanism and construction diversity, implementability, novelty relative to the existing library, and downstream admission under matched research budgets. Matching data, models, report availability, and evaluation periods is necessary to isolate initialization effects. The second stage of the present workflow inherits the first stage's resources and is therefore not the unguided condition.

\paragraph{Experience, breadth, and resource allocation.}
Experience may help research by transferring a useful construction, avoiding a known failure, or improving diagnostic choices. It may also preserve mistakes. Frozen-library and evolving-library comparisons can test these possibilities while recording access, adoption, and subsequent actions. The output cohorts describe discoveries under changing configurations. Complete attempt histories, library size at task launch, and actual resource use are needed to estimate diminishing returns or saturation; the 70-task episode in Appendix~\ref{sec:local-production} starts from the same formal library size throughout. Causal effects of width, concurrency, or their allocation against depth require matched resource studies. \emph{From Search to Research} provides related benchmark evidence on model capability and research budgets \citep{deng2026search}; this paper does not repeat those comparisons.

\paragraph{Production coverage and reproducibility.}
Production totals combine recovered output records with an author-supplied summary of the latest 133-output batch. The summary reports screening outcomes but does not provide individual factor artifacts for the common-snapshot evaluation. Complete task histories, whole-campaign runtime, invocation cost, and the mapping from historical deployment labels to underlying models remain unavailable.

\paragraph{Historical performance and generalization.}
The factor library was selected using historical information, and repeated backtesting can overfit the research period \citep{bailey2017overfitting}. Rolling combination weights constrain the timing of coefficient estimation but do not remove prior factor-selection bias. The portfolio uses approximate execution, constant fees, a price-index reference, and last-price valuation of residual holdings without fresh quotes. Reserved discovery-time evaluation and fuller execution accounting are needed to assess generalization and deployable performance.

\paragraph{From factors to autonomous strategy research.}
The same test-driven interface could be extended to portfolio construction, risk controls, trading rules, or the research process itself. This connects to broader AutoResearch and recursive self-improvement questions: an agent would propose strategy or process changes, evaluate them, and retain justified improvements. Such extensions require their own objectives, execution constraints, budgets, and held-out assessment. The simple factor portfolio in this paper is an application of the discovered signals, not an evaluation of autonomous strategy construction or recursive self-improvement.

%% file: sections/conclusion.tex
\section{Conclusion}
Auto-Alpha connects economic idea generation, executable factor research, and persistent research libraries. Report-guided initialization supplies both admitted signals and candidates that later proposals can develop. Continued research expands the range of constructions, while testing and similarity review determine which outputs enter the library. Factor diagnostics and a simple portfolio demonstrate the historical behavior of these retained artifacts. The resulting workflow provides a basis for studying how initialization, experience, and research budgets affect autonomous discovery under controlled evaluation.

%% file: sections/data_setup.tex
\section{Data Sources, Frequencies, and Timing}
\label{sec:data-setup}
The research environment combines market observations at several frequencies with corporate disclosures and reference data. Table~\ref{tab:data-setup} summarizes the supplied catalogue. Market, financial, and reference datasets are sourced from Tushare; the supplied daily style-exposure dataset is sourced from JoinQuant. These are available research inputs: each implementation requests the fields it needs, and no factor is assumed to use every source. Intraday inputs can describe within-day behavior, while quarterly disclosures can remain informative between announcements; both can contribute to the same daily stock ranking. The portfolio's weekly rebalancing is a downstream decision, separate from raw-data and signal frequency.

\begin{table}[htbp]
\centering\small
\renewcommand{\arraystretch}{1.1}
\caption{\textbf{Data available to the research environment.} Frequencies describe the source records, not a requirement to trade at that frequency. Financial-report periods and information-availability dates are distinct.}
\label{tab:data-setup}
\begin{tabularx}{\textwidth}{@{}>{\raggedright\arraybackslash}p{.20\textwidth}>{\raggedright\arraybackslash}p{.19\textwidth}>{\raggedright\arraybackslash}X@{}}
\toprule
Data family & Native frequency & Representative contents and research use \\
\midrule
Equity market data & Daily & Open, high, low, close, volume, traded amount, turnover, valuation ratios, capitalization, and adjustment factors; price, liquidity, and valuation signals. \\
Intraday market data & 1, 5, 15, 30, 60 minutes & OHLC, volume, and traded amount; within-day price paths, volatility, and trading activity. \\
Order-size money flows & Daily & Buy and sell volume and amount for small, medium, large, and extra-large orders; conditional flow signals. \\
Financial statements & Quarterly reporting; disclosure-driven updates & Income statements, balance sheets, and cash-flow statements, including revenue, assets, profits, and operating cash flows. \\
Earnings disclosures & Irregular announcements & Earnings forecasts and preliminary earnings reports; surprise measures and earlier updates to financial information. \\
Industry and style data & Effective-date intervals; daily exposures & Shenwan industry histories and CNE5-style exposures; economic grouping, exposure diagnostics, and configured controls. \\
Universe and index data & Daily state/prices; monthly index weights & Listing age, ST status, trading calendar, index prices, and constituent-weight snapshots; eligibility, execution context, and benchmarks. \\
\bottomrule
\end{tabularx}
\end{table}

\subsection{Information Available at the Signal Date}
For signal day $t$, the loader supplies daily and intraday observations only through $t$ within the requested lookback. Accounting records are filtered on the configured publication field: \texttt{f\_ann\_date} for income, balance-sheet, and cash-flow statements, and \texttt{ann\_date} for earnings forecasts and preliminary reports. Only records with that date no later than $t$ are exposed. The convention is an end-of-day calculation that includes that day's disclosures, followed by execution on a later trading day. The source fields encode dates rather than precise intraday publication timestamps. Multiple revisions of the same reporting period can remain visible; selecting the latest available revision and enforcing staleness limits are implementation responsibilities. Appendix~\ref{sec:case-implementation} illustrates these decisions with actual factor code.

Raw daily and intraday price fields are stored with adjustment factors; the loader applies the catalogue's specified adjustments. Their units also differ: daily volume is recorded in 100-share lots and traded amount in thousands of yuan, whereas minute volume and amount are in shares and yuan. These units must be aligned when constructing cross-frequency quantities. Daily rankings use the eligible stock set with available observations. The default loader requires a valid daily close, at least 60 calendar days since listing, and non-ST status, and excludes Beijing-listed stocks. Price-limit and suspension constraints on orders are handled by the execution layer rather than by removing every affected signal observation. The portfolio is not restricted to CSI 1000 constituents.

\subsection{Coverage and Evaluation Snapshots}
The inspected daily-equity file and all five minute-frequency directories extend from 2009-01-05 to 2026-11-03. Daily money-flow and style-exposure file bounds extend from 2010-01-04 to 2026-11-03. These are observed file or partition bounds, not claims of complete coverage for every stock, field, or bar; the minute directories differ in their date counts. Financial inputs have different disclosure histories and update asynchronously. The saved library evaluations summarized in the main text span 2013-01-04 through 2026-11-03, subject to each factor's data requirements and forward-label availability. The portfolio runs over 2020-01-02--2026-11-03, with earlier factor observations used to initialize rolling combination weights. The six archived versions in Appendix~\ref{sec:real-case} instead use evaluations ending on 2026-07-17. Their iteration metrics are kept separate from the later library snapshot.

The CSI 1000 benchmark uses daily price-index returns. Its local source file lacked eight trading dates between 2026-08-13 and 2026-08-24. For this study, the missing observations were retrieved from the same Tushare source into a separate benchmark snapshot; seven overlapping close observations agreed within $10^{-4}$. No interpolation or return filling was used. This snapshot supplies the benchmark comparison in Section~\ref{sec:empirical}.

\subsection{Supplementary Return Horizons}
\label{sec:supplementary-horizons}
The main text uses 1-day Rank IC and ICIR. Table~\ref{tab:library-horizons} retains 5- and 10-day results as supplementary descriptions of the same admitted library, with the forward-return definition in Section~\ref{sec:empirical}. Longer-horizon labels overlap across adjacent signal dates. Their unannualized ICIR values describe temporal stability and are not independent-observation significance tests. These additional horizons do not determine the representative-factor ordering or the main type-distribution plot.

\input{tables/library_horizon_summary}

%% file: tables/library_horizon_summary.tex
\begin{table}[htbp]
\centering\small
\caption{\textbf{Supplementary diagnostics across return horizons.} One day is the primary horizon; five and ten days are supplementary. Snapshot evaluated over 2013-01-04--2026-11-03. IC is daily cross-sectional Spearman correlation with the indicated forward VWAP return; ICIR is its time-series mean divided by standard deviation, without annualization. Quantiles are across factors. These are retrospectively selected library artifacts.}
\label{tab:library-horizons}
\begin{tabular}{@{}lrrrrr@{}}
\toprule
Horizon & IC 25th pct. & Median IC & IC 75th pct. & Maximum IC & Median ICIR \\
\midrule
1 day & 0.0204 & 0.0301 & 0.0378 & 0.0651 & 0.530 \\
5 days & 0.0318 & 0.0472 & 0.0577 & 0.0945 & 0.821 \\
10 days & 0.0373 & 0.0556 & 0.0672 & 0.1059 & 0.938 \\
\bottomrule
\end{tabular}
\end{table}

%% file: sections/appendix.tex
\section{Case Selection and Interpretation}
\label{sec:provenance}
The qualitative findings use selected idea records, implemented versions, research notes, and experience entries. These cases illustrate decisions within the test-driven loop; they are not a census of either production stage and do not establish the prevalence of a behavior. Production counts instead use the cohort and version accounting in Appendix~\ref{sec:production-accounting}.

The improvement, plateau, and failure cases are \texttt{neglect\_\allowbreak ato\_\allowbreak improver}, \texttt{gp\_\allowbreak sue}, and \texttt{loss\_\allowbreak recog\_\allowbreak asym}. The experience-writeback case is \texttt{limit\_\allowbreak resilience\_\allowbreak rate}. Their numerical sequences come from saved research evaluations; auxiliary diagnostics are not counted as additional named versions. Appendix~\ref{sec:real-case} links an initial idea to implementation, feedback, and the delivered factor. These examples are not relabeled as stage-specific outcomes without matching their production records.

\subsection{Example Factors}
\label{sec:factor-examples}
Table~\ref{tab:factor-examples} illustrates individual signals from the evaluated library.
\input{tables/factor_examples}
\FloatBarrier

\section{Implementation Contract and Reuse}
The operational contract has three main interfaces: an idea and its source, a research environment with tools and allowed edits, and a candidate package for assessment. The idea generator persists structured JSON, while the current launch path primarily passes idea text and a source reference into the research session. The paper's structured example should therefore not be interpreted as a schema that is uniformly enforced at every stage. Likewise, data-timing conventions and evaluator-edit restrictions include prompt-level instructions and should not be mistaken for a security sandbox.

Finalization saves the selected candidate's artifacts and invokes a separate audit process when configured. Audit findings and diagnostic metrics inform admission, but the current promotion operation does not itself require proof that every review check passed. The architecture describes the intended human-confirmed workflow. Stronger automated gate enforcement, validated experience schemas, and exact budget accounting are possible engineering improvements; none is assumed to have been completed in this account.

Reusing the methodology requires making the research contract explicit in the new environment. The data catalogue must specify available fields and timing; the evaluator must define return construction, costs, tradability, and diagnostics; the agent must know which code and configuration it may change. Candidate records should retain the initial idea, selected implementation, configuration, source material, diagnostic history, and admission decision. Experience entries should additionally retain their originating evidence and applicability conditions. This contract makes research inspectable even when the choice of model, data source, or orchestration software changes.

%% file: tables/factor_examples.tex
\begin{table}[htbp]
\centering\small
\caption{\textbf{Individual factor examples from the formal library.} First six rows have the highest 1-day Rank IC in this snapshot; the last row is the traceable idea-to-factor case in Appendix~\ref{sec:real-case}. The composite portfolio uses the full factor library, not this displayed subset. Rank IC and ICIR use the one-day horizon and the period in Table~\ref{tab:library-ic}. Net LS Sharpe is the single-factor long--short diagnostic with 10bp one-way costs, distinct from portfolio-account Sharpe.}
\label{tab:factor-examples}
\begin{tabular}{@{}lrrr@{}}
\toprule
Factor identifier & Rank IC (1d) & ICIR (1d) & Net LS Sharpe \\
\midrule
{\ttfamily\footnotesize ewmavol\_dm\_semiup} & 0.0651 & 0.711 & 3.828 \\
{\ttfamily\footnotesize cpv3\_wb60} & 0.0566 & 0.906 & 3.834 \\
{\ttfamily\footnotesize trend\_fund\_herd\_v5\_thrpast} & 0.0518 & 0.771 & 4.035 \\
{\ttfamily\footnotesize asp\_v2\_rcs} & 0.0509 & 0.789 & 3.841 \\
{\ttfamily\footnotesize network\_centrality\_scc} & 0.0469 & 0.628 & 3.752 \\
{\ttfamily\footnotesize amplitude\_volume\_weighted} & 0.0467 & 0.813 & 4.687 \\
\midrule
{\ttfamily\footnotesize neglect\_ato\_improver\_v6} & 0.0378 & 0.538 & 3.525 \\
\bottomrule
\end{tabular}
\end{table}

%% file: sections/real_case.tex
\section{A Real Generated Idea and Its Delivered Factor}
\label{sec:real-case}
The example below comes from the saved idea and production run for \texttt{neglect\_\allowbreak ato\_\allowbreak improver}. The delivered artifact, \texttt{neglect\_\allowbreak ato\_\allowbreak improver\_\allowbreak v6}, is present in the formal Factor Library. The JSON is an abridged English translation of the original Chinese record, with the original 60-day proposal preserved. The source archive includes the original idea JSON and the delivered Python implementation under \texttt{examples/}.

\subsection{Step 1: Generate an Idea}
\label{sec:case-idea}
\noindent\begin{minipage}{\linewidth}
\begin{lstlisting}
{
  "name": "neglect_ato_improver",
  "idea": "Multiply the cross-sectional percentile rank of the year-on-year change in quarterly asset turnover by the rank of neglect, measured using about 60 days of mean turnover after removing the effect of float market capitalization. Use announcement-date accounting records, reconstruct quarterly revenue from YTD figures, and exclude financial and real-estate firms.",
  "rationale": "Improving operating efficiency may be incorporated into prices slowly for stocks receiving little market attention, as subsequent reports reveal the improvement.",
  "data_needed": "daily turnover and float market capitalization; revenue; total assets",
  "source": {
    "type": "direction",
    "ref": "Inspect existing factors for underexplored combinations of economic mechanisms, data frequencies, and constructions."
  }
}
\end{lstlisting}
\end{minipage}

\subsection{Step 2: Implement the Signal}
\label{sec:case-implementation}
The delivered factor combines an operating-efficiency component with a size-adjusted attention component. For stock $i$, let $q_i(t)$ be the latest quarter for which the required accounting values can be computed using disclosures available by day $t$. Define
\begin{align}
 A_{i,q} &= \frac{\mathrm{Revenue}^{\mathrm{quarter}}_{i,q}}
 {\tfrac12(\mathrm{Assets}_{i,q}+\mathrm{Assets}_{i,q-1})},\\
 \Delta A_{i,t} &= A_{i,q_i(t)}-A_{i,q_i(t)-4}.
\end{align}
Quarterly revenue is reconstructed from year-to-date disclosures, with the first quarter treated directly rather than differenced across years. The implementation uses income statements, balance sheets, and earnings express disclosures through the framework's as-of interface. It excludes financial firms and masks stale accounting observations beyond its five-quarter threshold.

For the attention component, the final code averages free-float turnover over the latest 10 trading days, requiring 10 positive valid observations. It clips the logarithm of that mean at the cross-sectional 1st and 99th percentiles, then regresses it on an intercept and log float market capitalization. Let $e_{i,t}$ be the residual. The raw factor is
\begin{equation}
 f^{\mathrm{raw}}_{i,t}
 =\operatorname{PctRank}_t(\Delta A_{i,t})\,
  \operatorname{PctRank}_t(-e_{i,t}),
\end{equation}
where each rank is computed on the available observations for that component and the product uses their common stock set. High values indicate both efficiency improvement and low turnover relative to firm size. The saved postprocessing pipeline applies MAD clipping, industry and size neutralization, and centered ranking. The core combination is directly visible in the delivered code:

\begin{lstlisting}
r_dato = dato.rank(pct=True)
r_neglect = self._neglect_leg(window)
common = r_dato.index.intersection(r_neglect.index)
score = r_dato.reindex(common) * r_neglect.reindex(common)
\end{lstlisting}

\subsection{Step 3: Interpret Feedback and Revise}
\label{sec:case-feedback}
The archived versions show the research agent making concrete choices within this idea. Tightening the accounting-staleness threshold produced little change. Shortening the attention window improved the recorded diagnostics, after which the agent added earnings express disclosures. Table~\ref{tab:case-versions} reports the one-day IC and ICIR from each version's saved metrics, alongside net Sharpe. The original research notes considered several diagnostics and horizons; this presentation does not imply that the agent selected versions solely by one-day IC. These version-level evaluations end on 2026-07-17 and are distinct from the current library evaluations ending on 2026-11-03. They are a record of adaptive historical search, not independent validation trials.

The feedback affected both what the agent changed and what it retained. At v1, the saved diagnostics include 1-day Rank IC 0.0279, net long--short Sharpe 3.189, and average daily turnover 3.60\%. Reducing the accounting-staleness threshold from five quarters to four barely changed the results (v2 Sharpe 3.198), so the agent retained five quarters. It then tested shorter attention windows. At v5, the 10-day window raised 1-day Rank IC to 0.0376 and net Sharpe to 3.577, while turnover rose to 7.56\%. The agent therefore inspected the gain after costs, rather than relying on IC alone, and next tested whether earlier earnings disclosures improved the accounting leg.

\input{tables/case_versions}

\subsection{Step 4: Select, Audit, and Retain the Artifact}
\label{sec:case-assessment}
The run selected v6 after adding preliminary earnings reports: saved 1-day Rank IC was 0.0377, net long--short Sharpe 3.605, and average daily turnover 7.61\%. Its largest absolute mean rank correlation with the comparison library was 0.444. The delivery contains the implementation, postprocessing configuration, evaluation metrics, research card, and audit record. The selected implementation is also present in the formal Factor Library. This is an observed admission outcome; it does not establish an automatic promotion rule or reconstruct the human review discussion.

The saved LLM audit returned a leakage verdict of \texttt{PASS} and an idea-alignment score of 4 out of 5. It explicitly noted the change from a roughly 60-day neglect measure to a 10-day measure. It also recorded that the initial idea's real-estate exclusion was not implemented as a hard filter; industry neutralization remained in the pipeline, which is not equivalent to exclusion. These findings illustrate the role of assessment in preserving a visible distinction between the proposal and the delivered implementation. The audit is a recorded model judgment, not a proof of point-in-time correctness or generalization.

The audit's two principal fields are reproduced below; its textual findings are summarized above. The source package includes the original audit JSON.
\begin{lstlisting}
{
  "leak": {"verdict": "PASS"},
  "idea_alignment": {"score": 4}
}
\end{lstlisting}

\subsection{Step 5: Extract a Conditional Research Lesson}
\label{sec:case-lesson}
A reusable lesson suggested by this trajectory is that a fast component used to condition a slow fundamental signal may require a different window from the same component used alone. A later investigation could retrieve that lesson and compare attention windows while tracking coverage and costs. The archive also contains auxiliary diagnostics attempting to separate window freshness from valid-observation filtering. Their interpretation remains conditional on this construction and evaluation setting; the trajectory does not establish a universal optimal attention window.

The preceding lesson is a synthesis of this run's notes. Appendix~\ref{sec:experience-examples} separately shows a verified library writeback and subsequent research's use of an existing entry; these examples come from different trajectories.

%% file: tables/case_versions.tex
\begin{table}[htbp]
\centering\small
\caption{\textbf{Six saved versions of the real case.} Archived evaluations over 2013-01-01--2026-07-17. Net Sharpe uses the single-factor long--short evaluation and its 10bp one-way costs; it is not the Top-300 account Sharpe. IC and ICIR use the one-day horizon extracted from each saved metrics file. These are adaptively selected versions, with additional diagnostics between them.}
\label{tab:case-versions}
\begin{tabularx}{\textwidth}{@{}lXrrr@{}}
\toprule
Version & Main change & IC (1d) & ICIR (1d) & Net Sharpe \\
\midrule
v1 & Original proposal; 60-day turnover window & 0.0279 & 0.400 & 3.189 \\
v2 & Staleness threshold: 5 to 4 quarters & 0.0279 & 0.400 & 3.198 \\
v3 & 40-day turnover window; full-window validity & 0.0298 & 0.421 & 3.280 \\
v4 & 20-day turnover window & 0.0340 & 0.483 & 3.514 \\
v5 & 10-day turnover window & 0.0376 & 0.545 & 3.577 \\
v6 & Add earnings express disclosures & 0.0377 & 0.547 & 3.605 \\
\bottomrule
\end{tabularx}
\end{table}

%% file: sections/experience_examples.tex
\section{Experience Retrieval and Writeback Examples}
\label{sec:experience-examples}
The Experience Library connects investigations through reusable research advice. The examples below supplement the main case with two other recorded runs: \texttt{gp\_sue} for retrieval and \texttt{limit\_resilience\_rate} for writeback. They are separate trajectories, not missing steps inferred to have occurred in the asset-turnover case. The examples establish recorded use and stored content; they do not isolate the library's causal contribution.

\subsection{Retrieve, Apply, and Check Applicability}
In \texttt{gp\_sue}, the agent read an entry recommending additional book-to-price and nonlinear-size controls for expansion-confirmation factors. The run notes explicitly cited this advice when proposing the second version. The agent kept the signal construction fixed and added those postprocessing controls. Recorded net long--short Sharpe increased from 2.954 to 3.345, and the corresponding saved 1-day Rank IC increased from 0.0082 to 0.0095. Six later construction variants did not surpass v2, which was retained. This is a concrete sequence of retrieved advice, an implementation choice, observed feedback, and candidate selection.

The asset-turnover case illustrates why the advice needs conditions. Its notes questioned whether an operating-efficiency improvement had the same exposure structure as an expansion-confirmation factor. Testing book-to-price controls alone changed net Sharpe from 3.189 to 3.220, while adding nonlinear-size controls as well reduced it to 3.106. The agent retained the original postprocessing configuration. These observations support selective testing of an entry's applicability; they do not establish that the entry's economic explanation is correct or that the same actions would not occur without retrieval.

\subsection{Write a Reusable Diagnostic Procedure}
The \texttt{limit\_resilience\_rate} run wrote an entry about distinguishing event frequency from conditional event magnitude in sparse-event factors. Its initial proposal compared recoveries from downward price-limit touches with failures to hold an upward price limit. The ratio version had recorded net Sharpe $-1.041$; a variant ranking only event stocks remained negative at $-1.104$. A negative event-frequency construction instead reached 1.490. The written entry recommends diagnosing these components before spending the next iterations on ratio smoothing or shrinkage. The following card is an abridged English rendering of that actual Markdown entry, reorganized for readability rather than presented as an enforced storage schema.

\begin{center}
\begin{minipage}{.96\textwidth}
\small
\setlength{\parskip}{4pt}
\textbf{Experience: separate sparse-event frequency from the event ratio.}\par
\textbf{When to retrieve it.} Most stocks have few or no events, and neutral-filled ratio values produce a large tied group.\par
\textbf{Next test.} Compare the original ratio, the ratio restricted to event stocks, and an event-count signal; inspect quantile shape, IC, costs, and exposure to the existing library.\par
\textbf{Recorded evidence.} In the originating run, the ratio variants were negative, while the event-frequency version had net Sharpe 1.490. Shortening the latter's window raised Sharpe only to 1.530 and worsened quantile monotonicity from 0.915 to 0.855, so the earlier version was retained.\par
\textbf{Scope.} Dense events or an already informative ratio can require a different investigation. Replacing a ratio with event frequency also changes the hypothesis and must be recognized in assessment. These observations are evidence from the originating setting, not a universal rule about event factors.
\end{minipage}
\end{center}

The reusable artifact is a diagnostic decision procedure with a source and applicability conditions. It retains more than the best parameter or final score: a later agent can test whether its own signal has the same tied-group behavior and compare the same competing explanations. The source package includes the original experience entry. The available records do not establish how often this particular entry was subsequently reused successfully.

%% file: sections/factor_diagnostics.tex
\section{Long--Short and Decile Diagnostics}
\label{sec:factor-diagnostics}
\begin{figure}[htbp]
\centering
\includegraphics[width=0.94\textwidth]{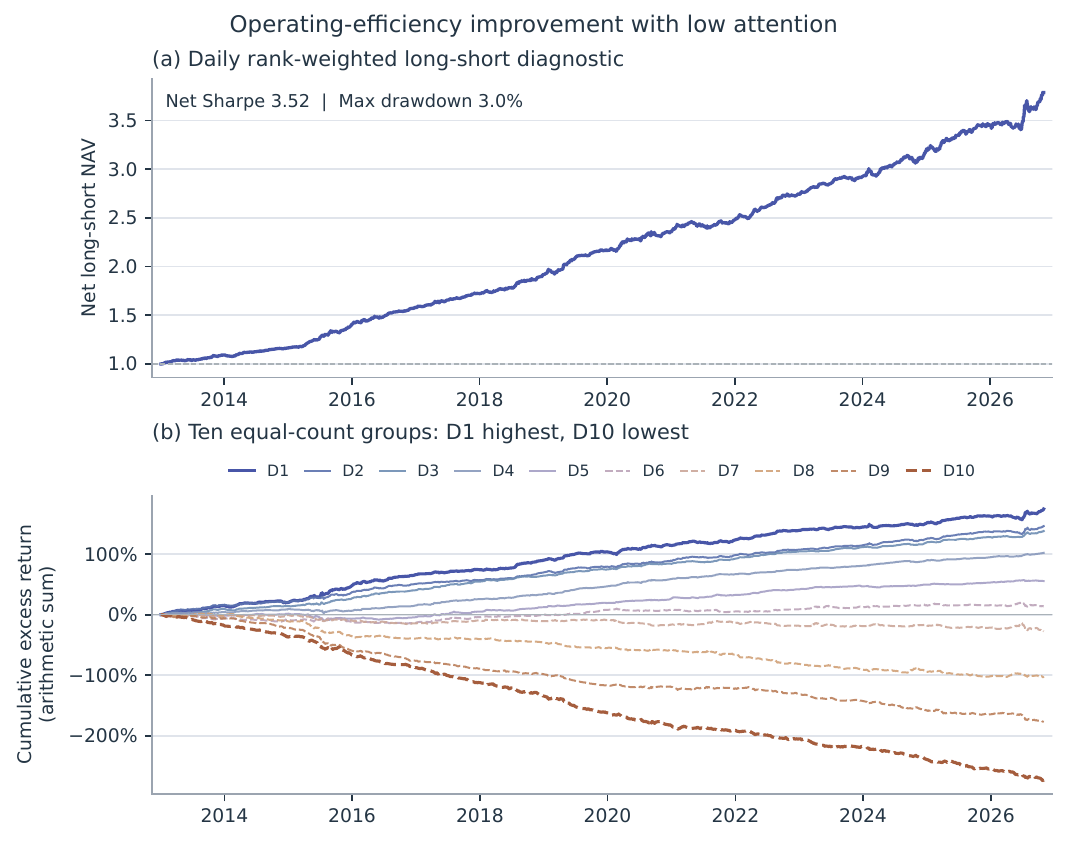}
\caption{\textbf{Diagnostics of the actual delivered factor.} The operating-efficiency improvement and low-attention factor described in Appendix~\ref{sec:real-case}. Top: net daily rank-weighted long--short NAV. Bottom: ten groups' cumulative excess returns relative to the eligible-pool equal-weight baseline; D1 has the highest scores. These are historical factor diagnostics rather than executable short-selling account returns.}
\label{fig:real-factor-diagnostics}
\end{figure}

\begin{figure}[htbp]
\centering
\includegraphics[width=\textwidth]{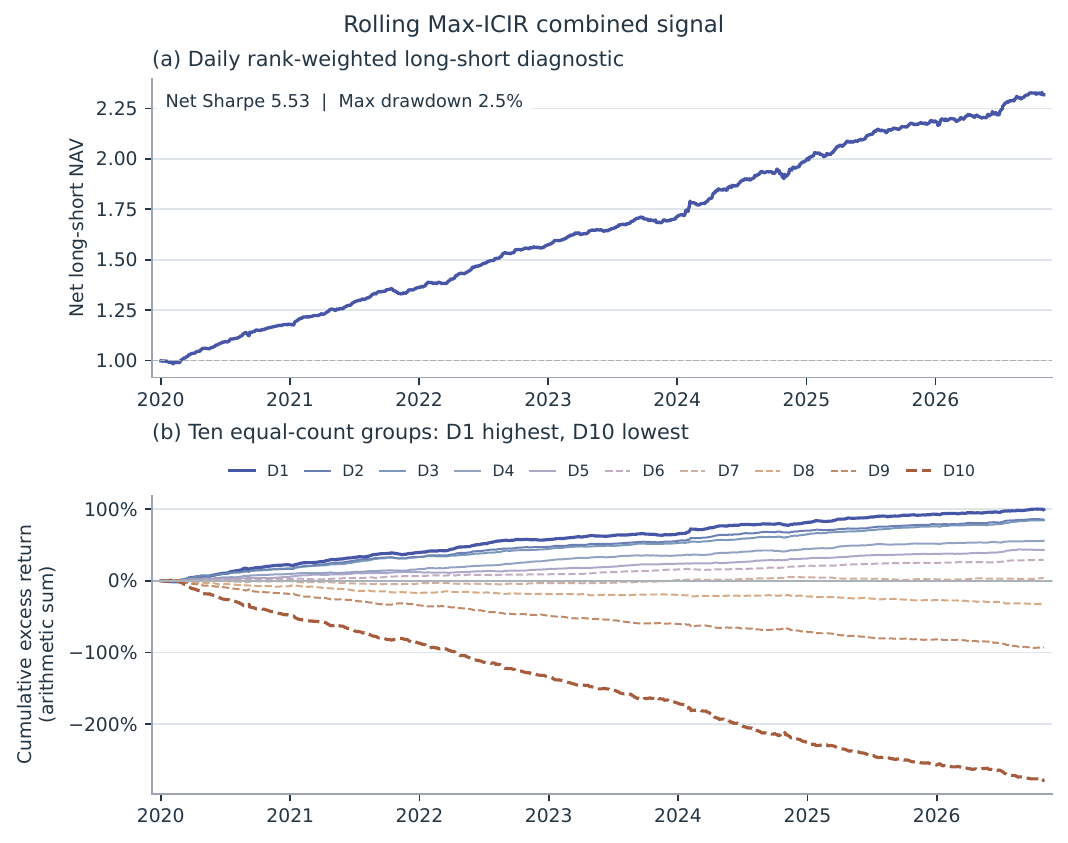}
\caption{\textbf{Diagnostics of the rolling Max-ICIR composite.} The same combined scores used for stock selection in the main account, evaluated here at a daily frequency over 2020--2026. Top: net rank-weighted long--short NAV. Bottom: ten groups' cumulative excess returns, using the same conventions as Figure~\ref{fig:real-factor-diagnostics}. Rolling factor weights use matured labels; the retrospective selection of the factor library remains a separate limitation.}
\label{fig:maxicir-diagnostics}
\end{figure}

Figures~\ref{fig:real-factor-diagnostics} and~\ref{fig:maxicir-diagnostics} evaluate the delivered factor over 2013--2026 and the Max-ICIR composite over 2020--2026. Both use saved signals and the original evaluator. These daily signal diagnostics are distinct from the weekly long-only account in Section~\ref{sec:empirical}.

Long--short weights are centered stock ranks, normalized to sum to zero with absolute weights summing to one (long and short exposure are each one-half). Net returns deduct $0.001\sum_i|w_{i,t}-w_{i,t-1}|$, starting from zero weights. Missing forward returns contribute zero without removing their weights, including the final unavailable-label sessions. Borrow availability is not modeled. Deciles are approximately equal-count groups, with D1 highest and D10 lowest. Their curves sum daily gross excess returns over the eligible-pool equal-weight baseline arithmetically; unlike the long--short curve, they do not compound or deduct costs.

\clearpage

%% file: sections/phase_quality.tex
\section{Stage-Specific Factor Composition and Quality}
\label{sec:phase-quality}
\begin{figure}[htbp]
\centering
\includegraphics[width=\textwidth]{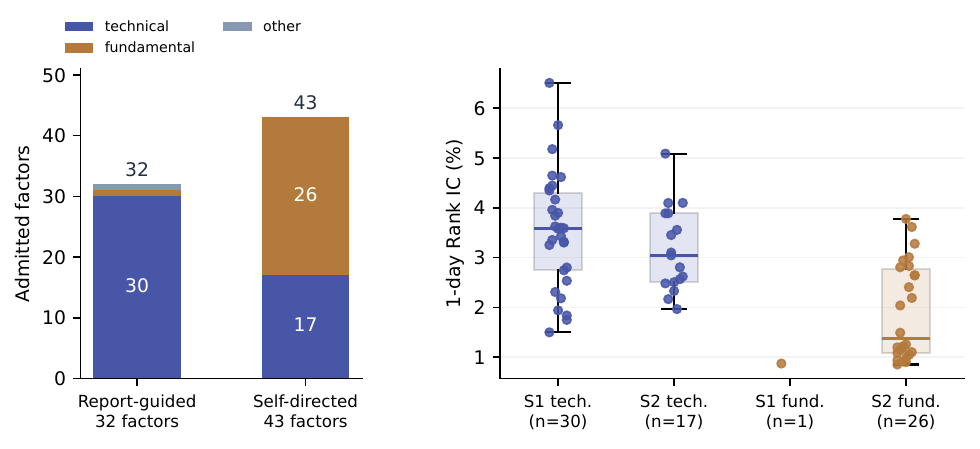}
\caption{\textbf{Factor composition and quality by research stage.} Left: counts by broad metadata category. Right: common-snapshot 1-day Rank IC. Stage 1 includes one ``other'' factor omitted from the within-type panel. These distributions describe admitted factors in the fixed evaluation snapshot.}
\label{fig:phase-quality}
\end{figure}

Figure~\ref{fig:phase-quality} compares 32 report-guided factors with 43 self-directed factors from the 105-factor evaluation snapshot. All use the common 2013-01-04--2026-11-03 evaluation and their admitted postprocessing configurations. Table~\ref{tab:phase-quality} summarizes quality within broad factor types.

\input{tables/phase_quality}

Self-directed research contributes more fundamental constructions, including fundamental--market interactions, while report-guided admissions are predominantly technical. Overall stage medians therefore combine composition differences with differences in research conditions. The single fundamental factor in Stage 1 does not support a stable within-type comparison.
\clearpage
\section{Recorded Outcomes of Candidate-Guided Refinement}
\label{sec:refinement-outcomes}
\begin{figure}[htbp]
\centering
\includegraphics[width=.92\textwidth]{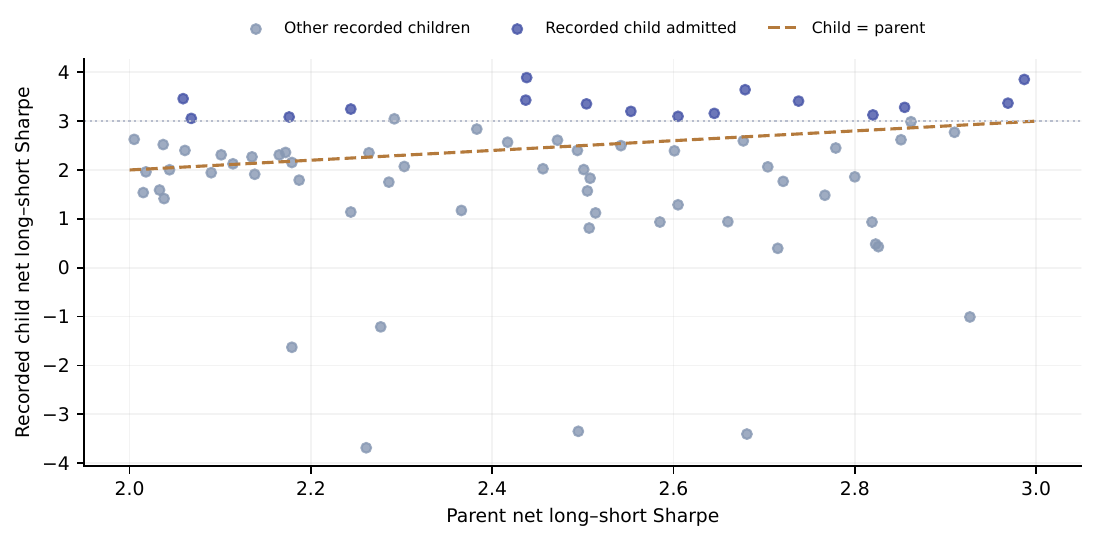}
\caption{\textbf{A proposed refinement does not guarantee a higher score.} Points are the 73 numerical comparisons in the refinement table. The diagonal marks equal parent and child Sharpe; the horizontal line is the admission performance threshold of 3. Highlighted points correspond to the recorded child names that are admitted. Subsequent same-idea revisions can change the selected version, so these highlights need not equal the 17 final admissions in the refinement cohort in Table~\ref{tab:production-funnels}.}
\label{fig:refinement-pairs}
\end{figure}

The refinement table contains 73 numerical parent--child comparisons drawn from 72 distinct parents. One parent has two proposals with numerical outcomes. Thirty scores improve, 43 decline, and the median difference is $-0.139$. Figure~\ref{fig:refinement-pairs} displays these recorded pairs. They are evidence about proposed revisions in this subset, not randomized comparisons or the marginal value of an extra iteration.

The paired comparison and the production funnel use different units. This appendix compares the parent and child versions recorded together; the funnel counts the final selected implementation from each linked proposal. The latter includes 81 implementations, 79 with metrics, and 17 final admissions. Later versions of the same idea do not create new discoveries, and proposals without a comparable parent evaluation do not enter Figure~\ref{fig:refinement-pairs}.

%% file: tables/phase_quality.tex
\begin{table}[htbp]
\centering\small
\caption{\textbf{Quality by research stage and factor type.} Values use the fixed 2013-01-04--2026-11-03 evaluation; ICIR is unannualized.}
\label{tab:phase-quality}
\begin{tabular}{@{}llrrr@{}}
\toprule
Stage & Type & $n$ & Median 1d IC & Median 1d ICIR \\
\midrule
1: report-guided & Technical & 30 & 0.0359 & 0.601 \\
1: report-guided & Fundamental & 1 & 0.0087 & 0.255 \\
1: report-guided & Other & 1 & 0.0365 & 0.528 \\
2: self-directed & Technical & 17 & 0.0305 & 0.564 \\
2: self-directed & Fundamental & 26 & 0.0137 & 0.374 \\
\bottomrule
\end{tabular}
\end{table}

%% file: sections/production_accounting.tex
\section{Production Units, Cohorts, and Admission Accounting}
\label{sec:production-accounting}
An idea specifies a proposed mechanism and construction; a research task develops it through one or more versions. We count selected or retained outputs after reconciling duplicate copies and explicitly linked versions. A new parent-based proposal is a separate investigation. A scored output has saved backtest metrics; qualification means recorded net long--short Sharpe at least 3. Screening, semantic audit, and human admission are separate decisions. Missing audit records are not failed audits.

\subsection{Report-Guided Ideas and Overlapping Screening Batches}
The 120 report-guided ideas produce 120 final implementations, with 37 qualified outputs and 31 final-version admissions. For \texttt{act\_upday\_lgmd}, an admitted v4 precedes the final v8, which is rejected for correlation with v4. This gives 32 admitted idea lineages without changing the final-output funnel.

An overlapping early screening summary reports 165 candidate outputs, 26 qualified names, and 21 admissions, overlapping with report-guided research. Thirteen qualified names and 11 admissions match report-guided final deliveries. The 21 admissions cannot all count again as Stage 2. Complete membership of the 165-output pool is unavailable, so it supplies no independent Stage 2 denominator. Stage assignment follows proposal source, not screening date.

\subsection{Continuous Mining and Candidate Refinement}
The directly linked cohorts contain seven outputs from an initial new-direction batch, 81 candidate-guided outputs, and 68 new-direction outputs from the timed production episode in Appendix~\ref{sec:local-production}. Links use version names, parent references, idea/run identifiers, and admission decisions. The refinement account corrects four outdated selected-version names and adds six scored and two code-only outputs to 73 numerical records. The original 73 paired scores remain unchanged in Appendix~\ref{sec:refinement-outcomes}.

The early cohort contains 127 historical scored outputs, excluding exact report-delivery aliases. We add ten admissions from the same research period whose metrics were later refreshed and whose origins are unrelated to reports, remove four identical code-and-metric copies, and merge two explicitly linked same-idea version pairs: $127+10-4-2=131$. Twelve qualify and ten are admitted. The authors confirm that these non-refinement outputs belong to continuous mining. Together with the 75 later new-direction outputs, they yield 206 continuous-mining outputs and 26 admissions before the latest batch.

The authors additionally report 133 continuous-mining outputs: 19 meet the Sharpe threshold, all 133 have maximum correlation below 0.7, and 19 pass both screens; the best Sharpe is 4.795. We use this batch summary without item-level reconstruction. Its 19 joint-screen passes are counted as selections, while formal admission status is not supplied. Combined totals are 339 continuous-mining outputs with 45 selections and 81 refinement outputs with 17 selections. The latest batch has no individual metrics in the evaluated snapshot.

Retained proposal files contain 171 distinct names: 93 candidate-guided and 78 direction-based. All 70 surviving run records belong to the timed production episode. These inventories differ in retention and coverage. An unlinked idea is unresolved, not failed; candidate counts and the timed episode do not establish whole-campaign attempts, runtime, or invocation cost.

\subsection{Screening Events and the Pooled Library}
Three successive screening batches contain 43, 12, and 32 qualified entries and admit 38, seven, and 21 factors, respectively. Five earlier rejections recur in each later batch: 87 screening events correspond to 77 distinct candidates and 66 admissions. The second screening batch uses a 40-factor seed; four separately approved fundamental factors enter before promotion, explaining the 44-factor inventory at execution.

The 105-factor evaluation population contains 32 report-guided lineages, 43 self-directed factors, and 30 prior or other factors. The initial two factors, 66 screening admissions, and four separate admissions yield 72. Two later direction-based and 17 candidate-guided additions yield the 91-factor library at late-task launch; 14 subsequent admissions yield 105.

Funnels use historical delivery or screening scores; stage quality uses the common evaluation snapshot ending on 2026-11-03 with admitted postprocessing. Packaged sanitized ledgers retain source-linkage methods, confirmed activity assignments, the latest batch summary, pairs, and screening counts, excluding raw reports and private conversations.

%% file: sections/local_production.tex
\section{A Timed Production Episode}
\label{sec:local-production}
A retained set of run records describes one timed production episode within the broader self-directed research campaign. They contain 70 starts and 69 finishes. One unfinished task and one finished task have no collected factor, and two collected factors lack saved metrics: 68 implementations are collected and 66 are scored. Eighteen meet the performance threshold, and 14 are admitted at subsequent screening. The first launch to the last recorded finish spans 39.8 hours (Table~\ref{tab:throughput}). This elapsed time covers the recorded episode rather than the entire two-stage campaign and does not measure compute cost.

\input{tables/research_throughput}

Configured concurrency changes from four to eight slots, and the research-session iteration budget changes from five to eight. These are prompt-level budgets rather than exact counts of backtest invocations. Historical deployment labels also vary, and their underlying model identities are not fully resolved in the retained records. We therefore do not treat the episode as a fixed-model or fixed-budget scaling experiment. Fourteen eventual admissions divided by 70 starts gives a retrospective episode fraction; the admissions occur later, not uniformly during the measured window.

\begin{figure}[htbp]
\centering
\includegraphics[width=\textwidth]{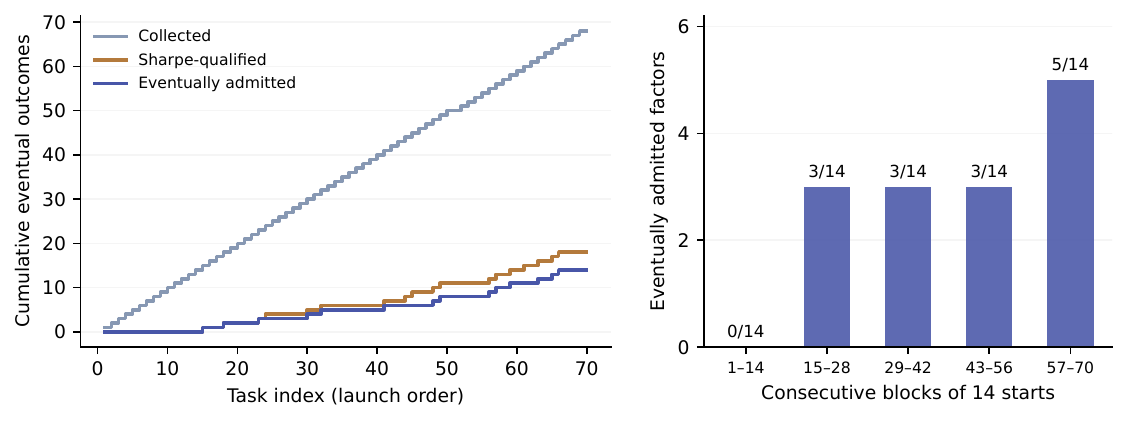}
\caption{\textbf{Local continued production, not a marginal-effect estimate.} Left: eventual outcomes ordered by task launch, rather than actual admission time. Right: eventual admissions in non-overlapping blocks of 14 starts, retaining the unsuccessful and unfinished tasks in this episode's denominator. The 70 records cover this episode only. Changing model configurations and budgets prevent a causal interpretation of the sequence.}
\label{fig:search-breadth}
\end{figure}

The five blocks yield zero, three, three, three, and five admissions. This establishes that later proposals in this episode still produce accepted factors; it neither demonstrates diminishing returns nor establishes an improving yield. Every task records the same 91-factor formal library at launch, so the sequence cannot estimate how discovery yield changes with formal library size. A whole-campaign saturation analysis requires complete attempt histories, task-start library snapshots, actual resource use, and resolved admission outcomes.

%% file: tables/research_throughput.tex
\begin{table}[htbp]
\centering\small
\caption{\textbf{A bounded production episode in Stage 2.} The measured research window spans the first launch to the last recorded finish. Admissions are later screening outcomes, not events occurring uniformly within the window. Configurations vary during the episode.}
\label{tab:throughput}
\begin{tabular}{@{}lr@{}}
\toprule
Measurement & Value \\
\midrule
Started / finished tasks & 70 / 69 \\
Collected / evaluated outputs & 68 / 66 \\
Sharpe-qualified / eventually admitted & 18 / 14 \\
Elapsed research window & 39.8 hours \\
Evaluated outputs per calendar hour & 1.66 \\
Eventually admitted / started tasks & 14/70 (20.0\%) \\
Configured concurrent slots & 4 to 8 \\
Research-session iteration budget & 5 to 8 \\
\bottomrule
\end{tabular}
\end{table}

%% file: references.bib
@article{deng2026search,
  title = {From Search to Research: Exploring Search Scaling in Autonomous Quantitative Factor Mining},
  author = {Deng, Kangcheng and Cai, Hui and Lu, Jiacheng and Qian, Chester Zhongshu and Sun, Rui and Luan, Beidi and Li, Jing and Jiang, Daxin and Bai, Zuo},
  journal = {arXiv preprint arXiv:2609.35559},
  year = {2026},
  url = {https://arxiv.org/abs/2609.35559}
}

@article{liu2026xalpha,
  title = {{X-Alpha}: An {AI} Quant Researcher with Memory-Driven Self-Evolution},
  author = {Liu, Fengyuan and Fu, Yuchen and Wang, Yuqi and others},
  journal = {arXiv preprint arXiv:2607.08332},
  year = {2026},
  note = {Version 3},
  url = {https://arxiv.org/abs/2607.08332}
}

@inproceedings{liu2026cogalpha,
  title = {Cognitive Alpha Mining via {LLM}-Driven Code-Based Evolution},
  author = {Liu, Fengyuan and Huang, Yi and Luo, Sichun and Wang, Yuqi and Yang, Yazheng and Li, Xinye and Hu, Zefa and Feng, Junlan and Liu, Qi},
  booktitle = {Proceedings of the 64th Annual Meeting of the Association for Computational Linguistics (Volume 1: Long Papers)},
  pages = {11715--11749},
  year = {2026},
  url = {https://aclanthology.org/2026.acl-long.538/}
}

@article{li2025rdagent,
  title = {{R\&D-Agent-Quant}: A Multi-Agent Framework for Data-Centric Factors and Model Joint Optimization},
  author = {Li, Yuante and Yang, Xu and Yang, Xiao and Xu, Minrui and Wang, Xisen and Liu, Weiqing and Bian, Jiang},
  journal = {arXiv preprint arXiv:2505.15155},
  year = {2025},
  url = {https://arxiv.org/abs/2505.15155}
}

@article{tang2025alphaagent,
  title = {{AlphaAgent}: {LLM}-Driven Alpha Mining with Regularized Exploration to Counteract Alpha Decay},
  author = {Tang, Ziyi and Chen, Zechuan and Yang, Jiarui and Mai, Jiayao and Zheng, Yongsen and Wang, Keze and Chen, Jinrui and Lin, Liang},
  journal = {arXiv preprint arXiv:2502.16789},
  year = {2025},
  url = {https://arxiv.org/abs/2502.16789}
}

@article{zhang2020autoalpha,
  title = {{AutoAlpha}: an Efficient Hierarchical Evolutionary Algorithm for Mining Alpha Factors in Quantitative Investment},
  author = {Zhang, Tianping and Li, Yuanqi and Jin, Yifei and Li, Jian},
  journal = {arXiv preprint arXiv:2002.08245},
  year = {2020},
  url = {https://arxiv.org/abs/2002.08245}
}

@article{shi2024alphaforge,
  title = {{AlphaForge}: A Framework to Mine and Dynamically Combine Formulaic Alpha Factors},
  author = {Shi, Hao and Song, Weili and Zhang, Xinting and Shi, Jiahe and Luo, Cuicui and Ao, Xiang and Arian, Hamid and Seco, Luis},
  journal = {arXiv preprint arXiv:2406.18394},
  year = {2024},
  url = {https://arxiv.org/abs/2406.18394}
}

@article{yuan2024alphagpt,
  title = {{Alpha-GPT 2.0}: Human-in-the-Loop {AI} for Quantitative Investment},
  author = {Yuan, Hang and Wang, Saizhuo and Guo, Jian},
  journal = {arXiv preprint arXiv:2402.09746},
  year = {2024},
  url = {https://arxiv.org/abs/2402.09746}
}

@article{wang2026factorminer,
  title = {{FactorMiner}: A Self-Evolving Agent with Skills and Experience Memory for Financial Alpha Discovery},
  author = {Wang, Yanlong and Xu, Jian and Zhang, Hongkang and Huang, Shao-Lun and Sun, Danny Dongning and Zhang, Xiao-Ping},
  journal = {arXiv preprint arXiv:2602.14670},
  year = {2026},
  url = {https://arxiv.org/abs/2602.14670}
}

@article{shinn2023reflexion,
  title = {{Reflexion}: Language Agents with Verbal Reinforcement Learning},
  author = {Shinn, Noah and Cassano, Federico and Berman, Edward and Gopinath, Ashwin and Narasimhan, Karthik and Yao, Shunyu},
  journal = {arXiv preprint arXiv:2303.11366},
  year = {2023},
  url = {https://arxiv.org/abs/2303.11366}
}

@article{lu2024aiscientist,
  title = {The {AI} Scientist: Towards Fully Automated Open-Ended Scientific Discovery},
  author = {Lu, Chris and Lu, Cong and Lange, Robert Tjarko and Foerster, Jakob and Clune, Jeff and Ha, David},
  journal = {arXiv preprint arXiv:2408.06292},
  year = {2024},
  url = {https://arxiv.org/abs/2408.06292}
}

@article{brown2024monkeys,
  title = {Large Language Monkeys: Scaling Inference Compute with Repeated Sampling},
  author = {Brown, Bradley and Juravsky, Jordan and Ehrlich, Ryan and Clark, Ronald and Le, Quoc V. and R{\'e}, Christopher and Mirhoseini, Azalia},
  journal = {arXiv preprint arXiv:2407.21787},
  year = {2024},
  url = {https://arxiv.org/abs/2407.21787}
}

@article{zhu2025agentscaling,
  title = {Scaling Test-time Compute for {LLM} Agents},
  author = {Zhu, King and Li, Hanhao and Wu, Siwei and Xing, Tianshun and Ma, Dehua and Tang, Xiangru and Liu, Minghao and Yang, Jian and Liu, Jiaheng and Jiang, Yuchen Eleanor and Zhang, Changwang and Lin, Chenghua and Wang, Jun and Zhang, Ge and Zhou, Wangchunshu},
  journal = {arXiv preprint arXiv:2506.12928},
  year = {2025},
  url = {https://arxiv.org/abs/2506.12928}
}

@article{garikaparthi2026researchgym,
  title = {{ResearchGym}: Evaluating Language Model Agents on Real-World {AI} Research},
  author = {Garikaparthi, Aniketh and Patwardhan, Manasi and Cohan, Arman},
  journal = {arXiv preprint arXiv:2602.15112},
  year = {2026},
  url = {https://arxiv.org/abs/2602.15112}
}

@article{rank2026posttrainbench,
  title = {{PostTrainBench}: Can {LLM} Agents Automate {LLM} Post-Training?},
  author = {Rank, Ben and Bhatnagar, Hardik and Prabhu, Ameya and Eisenberg, Shira and Nguyen, Karina and Bethge, Matthias and Andriushchenko, Maksym},
  journal = {arXiv preprint arXiv:2603.08640},
  year = {2026},
  url = {https://arxiv.org/abs/2603.08640}
}

@article{starace2025paperbench,
  title = {{PaperBench}: Evaluating {AI}'s Ability to Replicate {AI} Research},
  author = {Starace, Giulio and Jaffe, Oliver and Sherburn, Dane and Aung, James and Chan, Jun Shern and Maksin, Leon and Dias, Rachel and Mays, Evan and Kinsella, Benjamin and Thompson, Wyatt and Heidecke, Johannes and Glaese, Amelia and Patwardhan, Tejal},
  journal = {arXiv preprint arXiv:2504.01848},
  year = {2025},
  url = {https://arxiv.org/abs/2504.01848}
}

@article{zhang2026openfingym,
  title = {{OpenFinGym}: A Verifiable Multi-Task Gym Environment for Evaluating Quant Agents},
  author = {Zhang, Kaicheng and Ge, Wen and Jiang, Lei and Yang, Weixin and Langham-Lopez, Jordan and Yu, Jialin and Szpruch, Lukasz and Ni, Hao},
  journal = {arXiv preprint arXiv:2606.26350},
  year = {2026},
  url = {https://arxiv.org/abs/2606.26350}
}

@inproceedings{yu2023alphagen,
  title = {Generating Synergistic Formulaic Alpha Collections via Reinforcement Learning},
  author = {Yu, Shuo and Xue, Hongyan and Ao, Xiang and Pan, Feiyang and He, Jia and Tu, Dandan and He, Qing},
  booktitle = {Proceedings of the 29th ACM SIGKDD Conference on Knowledge Discovery and Data Mining},
  year = {2023},
  pages = {5476--5486},
  doi = {10.1145/3580305.3599831},
  url = {https://arxiv.org/abs/2306.12964}
}

@inproceedings{yao2023react,
  title = {{ReAct}: Synergizing Reasoning and Acting in Language Models},
  author = {Yao, Shunyu and Zhao, Jeffrey and Yu, Dian and Du, Nan and Shafran, Izhak and Narasimhan, Karthik and Cao, Yuan},
  booktitle = {International Conference on Learning Representations},
  year = {2023},
  url = {https://openreview.net/forum?id=WE_vluYUL-X}
}

@inproceedings{snell2025scaling,
  title = {Scaling {LLM} Test-Time Compute Optimally Can be More Effective than Scaling Parameters for Reasoning},
  author = {Snell, Charlie and Lee, Jaehoon and Xu, Kelvin and Kumar, Aviral},
  booktitle = {International Conference on Learning Representations},
  year = {2025},
  url = {https://proceedings.iclr.cc/paper_files/paper/2025/hash/1b623663fd9b874366f3ce019fdfdd44-Abstract-Conference.html}
}

@inproceedings{luo2026alphabench,
  title = {{AlphaBench}: Benchmarking Large Language Models in Formulaic Alpha Factor Mining},
  author = {Luo, Haochen and Ko, HoTin and Chen, Jiandong and Sun, David and Zhang, Yuan and Liu, Chen},
  booktitle = {International Conference on Learning Representations},
  year = {2026},
  url = {https://proceedings.iclr.cc/paper_files/paper/2026/hash/4f3820576130a8f796ddbf204c841487-Abstract-Conference.html}
}

@inproceedings{luo2026alphaqt,
  title = {{AlphaQT-Bench}: Diagnosing the Gap between Financial Code Generation and Quantitative Reasoning in {LLMs}},
  author = {Luo, Sichun and Huang, Yi and Meng, Shichang and Liu, Fengyuan and Li, Mukai and Yao, Qinghua and Hu, Zefa and Feng, Junlan and Liu, Qi},
  booktitle = {Findings of the Association for Computational Linguistics: ACL 2026},
  year = {2026},
  pages = {2871--2886},
  doi = {10.18653/v1/2026.findings-acl.138},
  url = {https://aclanthology.org/2026.findings-acl.138/}
}

@inproceedings{wijk2025rebench,
  title = {{RE-Bench}: Evaluating Frontier {AI} {R\&D} Capabilities of Language Model Agents against Human Experts},
  author = {Wijk, Hjalmar and Lin, Tao Roa and Becker, Joel and Jawhar, Sami and Parikh, Neev and Broadley, Thomas and Chan, Lawrence and Chen, Michael and Clymer, Joshua M and Dhyani, Jai and Ericheva, Elena and Garcia, Katharyn and Goodrich, Brian and Jurkovic, Nikola and Kinniment, Megan and Lajko, Aron and Nix, Seraphina and Sato, Lucas Jun Koba and Saunders, William and Taran, Maksym and West, Ben and Barnes, Elizabeth},
  booktitle = {Proceedings of the 42nd International Conference on Machine Learning},
  series = {Proceedings of Machine Learning Research},
  volume = {267},
  pages = {66772--66832},
  year = {2025},
  url = {https://proceedings.mlr.press/v267/wijk25a.html}
}

@article{bailey2017overfitting,
  title = {The Probability of Backtest Overfitting},
  author = {Bailey, David H. and Borwein, Jonathan M. and L{\'o}pez de Prado, Marcos and Zhu, Qiji Jim},
  journal = {Journal of Computational Finance},
  volume = {20},
  number = {4},
  year = {2017},
  doi = {10.21314/JCF.2016.322},
  url = {https://scholarworks.wmich.edu/math_pubs/42/}
}
